\documentclass[prb,twocolumn,aps,10pt]{revtex4-1}
\usepackage[colorlinks,bookmarks=false,citecolor=blue,linkcolor=blue,urlcolor=black]{hyperref}
\usepackage{amssymb,amsmath} 
\usepackage{amsfonts}
\usepackage{color}
\usepackage{braket}
\usepackage[utf8]{inputenc}
\usepackage[T1]{fontenc}
\usepackage{lmodern}
\usepackage{epsfig}
\usepackage{ulem}
\usepackage{blindtext}
\usepackage{cleveref}
\begin{document}

\title{Extended Bose-Hubbard Model with Dipolar and Contact Interactions}

\author{Krzysztof Biedro\'n$^1$, Mateusz \L{}\k{a}cki$^1$, and Jakub Zakrzewski$^{1,2}$}
\affiliation{\mbox{$^1$Instytut Fizyki imienia Mariana Smoluchowskiego, Uniwersytet Jagiello\'nski, \L{}ojasiewicza 11, 30-048 Krak\'ow, Poland}
\mbox{$^2$Mark Kac Complex Systems Research Center, Jagiellonian University, \L{}ojasiewicza 11, 30-348 Krak\'ow, Poland}
}

\begin{abstract}
  We study the phase diagram of the one-dimensional boson gas trapped inside an optical lattice with contact and dipolar interaction taking into account next-nearest terms for both tunneling and interaction.
  Using the density matrix renormalization group, we calculate how the locations of phase transitions change with increasing dipolar interaction strength for average density $\rho = 1$.
  Furthermore, we show an emergence of pair-correlated phases for a large dipolar interaction strength and $\rho \geq 2$, including a~supersolid phase with an incommensurate density wave ordering manifesting the corresponding spontaneous breaking of the translational symmetry.
\end{abstract}
\date{\today}

\maketitle

\section{Introduction}

	Ultracold gases loaded in optical lattices enable simulation of a~broad range of lattice gas models, most prominently the~Bose-Hubbard (BH) model~\cite{Jaksch1998} with Mott insulator (MI) to superfluid (SF) quantum phase transition~\cite{Greiner2002}. A~precise control of model parameters is achieved by optical potential manipulation or by advanced techniques such as Feshbach resonances~\cite{Chin2010,Frisch2015}. Long range dipolar interparticle interactions are often taken into account by adding a simple nearest-neigbour interaction term resulting in Extended Bose-Hubbard model (EBH), which has been a~topic of numerous theoretical ~\cite{DallaTorre06,Sinha2007,Lahaye2009,Ruhman12,Baranov2012,Wall2013,Gallemi13,Gammelmark13,Dutta2015,Gallemi16,Kawaki17,Cartarius17} and experimental~\cite{Baier2016} works.  
	
    A feature of ultracold gases is the ability to control the geometry of the underlying optical lattice potential or even possibility for implementation of a more complex unit cell. The boundary conditions of the potentials can be set by an external harmonic or a box trap, leading then to Open Boundary Conditions (OBC)~\cite{Zakrzewski2009,Gaunt2013} or by arranging a~system into a~ring-like or a~cylinder-like geometry~\cite{Lacki2016,Kim2018}, thus implementing Periodic Boundary Conditions (PBC). Notably, one-dimensional systems offer a~possibility for efficient many-body numerical simulations of the resulting lattice models by a~family of methods related to the Density Matrix Renormalization group (DMRG)~\cite{Schollwock2011,ITensor}.
	
	For one dimensional lattices the EBH model features not only MI and SF phases but also an isolator density wave (DW) characterized by infinite-range spatial order, topologically-protected Haldane Insulator (HI) with a nonzero value of a string order parameter and supersolid (SS) phases which show both spatial ordering and superfluid behavior~\cite{Rossini2012, Batrouni2014,Gremaud16}. 
	It has also been suggested that at the mean density $\rho=3/2$ the EBH model features Fibonacci anyon excitations \cite{Wikberg12,Duric17} corresponding to fractional domain walls between different DW phases.
	In this context, the mean field analysis \cite{Wikberg12} predicted the existence of SS phase between DW and SF phases in contrast to DMRG calculation\cite{Duric17}.
	
 The necessary strength of the dipole-dipole interactions is achieved for isotopes of dysprosium and erbium \cite{Lu2011,Aikawa2012}, Feshbach molecules~\cite{Chotia2012} or polar molecules~\cite{Sowinski2012,Moses2015,Moses2017}. More exotic phases such as checkerboard or stripe-ordered phases are possible for higher dimensional lattices~\cite{Sengupta2005,Batrouni2006,Mishra2009,Capogrosso-Sansone2010,Zaletel2014,Lewenstein2012} -- for a review see \cite{Dutta2015}.

	The BH/EBH models are motivated by an expansion of the field operators in the discrete basis defined by Wannier functions~\cite{Kohn1959,Jaksch1998} for the optical potential, followed by truncating the physics to the lowest Bloch band and neglecting hopping beyond the nearest-neighbors. The BH model includes then on-site interactions only while the EBH contains also density-density interactions on the nearest neighbor sites. The rigor of this procedure has been a~topic of extended research in the presence of fast time dependence \cite{Lacki2013a,Pichler2013}, or strong inter-atom interactions manifesting as the so-called density-dependent tunnelings~\cite{Sowinski2012,Maik2013,Dutta2015}, or even as a renormalization of model parameters due to a virtual population of higher bands \cite{Luhmann2012,Bissbort2012,Lacki2013}. Moreover, the coupling beyond the nearest neighbor has been included in studies which treated shallow optical lattices~\cite{Kollath2010,Trotzky2012} or for strongly interacting dipolar 
systems~\cite{Lewenstein2002}. In the latter case the 
extra couplings led to an appearance of spatially ordered phases~\cite{Sengupta2005}. 

	Extensive studies of the EBH-like models mentioned in this section were mostly done by scanning the parameter space of the constructed Hamiltonians at a chosen mean density or possibly other constraints such as a ratio between parameters. In this study we take a more systematic approach to obtain Hamiltonian for a dipolar gas of ultracold atoms in the optical lattice and study its phase diagram. First, out intent is to modify only experimentally accessible parameters such as the optical lattice potential depth, scattering length for contact interactions, the dipole-dipole interaction strength and the mean density of gas. Second, we chose to keep all the relevant tight binding terms describing tunneling and interactions. In this way the parameters of the obtained EBH-like Hamiltonians yield a realizable physical model. In other words we get a natural constraints values of parameters.  This saves us from considering parameter ranges unaccessible experimentally. In the phase diagram defined by the 
experiment-like control knobs, we predict modifications of up to date theoretical results going beyond a simple readjustment of phase boundaries. In particular we provide evidence for the emergence of new phase --- a pair superfluid phase with an incommensurate density wave order.

In Section \ref{sec:mod} we derive the model from the microscopic principles identifying realistic parameter set relevant for ultracold dipolar atoms and ultracold dipolar molecules.
The phase diagrams for the system are presented in Section \ref{sec:rh1} (for the case of unit density in the lattice) and Section \ref{sec:rhv} (for the case of other densities)a.
In section \ref{sec:con} we provide final conclusions and outlooks. 
We finish with three appendices describing in detail the computational methods used throughout the paper: in Appendix \ref{app:wan} we present our method of calculating the terms present in the Hamiltonian, Appendix \ref{app:dmrg} contains the parameters used in our DMRG runs and in Appendix \ref{app:ssd} we describe the DMRG method used in Section \ref{sec:rhv}.

\section{Model}\label{sec:mod}

The realistic Hamiltonian that models ultracold bosonic gas in the one-dimensional optical lattice potential considered in this work will be of the form:
\begin{equation}
\label{modelHam}
\begin{split}
  H =& -t \sum_{i=1}^{L-1} \left( b^{\dag}_i b_{i+1} + \text{h.c.} \right) - t_{\textrm{nnn}} \sum_{i=1}^{L-2} \left( b^{\dag}_i b_{i+2} + \text{h.c.} \right) \\
   & + \frac{U}{2} \sum_{i=1}^{L} n_i(n_i - 1) + V \sum_{i=1}^{L-1} n_i n_{i+1} + V_{\textrm{nnn}} \sum_{i=1}^{L-2} n_i n_{i+2} \\
   & - T \sum_{i=1}^{L-1} \left[ b^{\dag}_i (n_i + n_{i+1}) b_{i+1} + \text{h.c.} \right],
\end{split}
\end{equation}
where $t$, $T$  and $V$ denote the amplitude for standard, nearest neighbor tunnelings, the amplitude of density-dependent tunnelings resulting from interactions and the strength of interactions between nearest neighbor sites, respectively. The terms proportional to $t_{\textrm{nnn}}$ and $V_{\textrm{nnn}}$ are respectively the tunneling and strength of interaction between next-nearest neighbor lattice sites.

The Hamiltonian \eqref{modelHam} in its full glory is a result of a realistic tight-binding approximation to the many-body formulation continuous in space, as given by the second quantization.
We consider an ultracold gas of atoms or molecules of mass $m$ in the separable optical potential created by three pairs of standing waves of lasers with a~wavelength $\lambda_L$ which takes the form $V_{\textrm{opt}}(\mathbf{r}) = V_x \cos^2(k_Lx) + V_y \cos^2(k_Ly) + V_z \cos^2(k_Lz)$ with  $k_L=2\pi/\lambda_L$.
The recoil energy $E_R = {\hbar^2 k_L^2}/{2 m}$ defines a~natural energy scale for the single-particle physics. We take $V_y = V_z = 50 E_R$ and $V_x\ll V_y,V_z$ which freezes the motion in directions $y, z$ and leaves an effectively one dimensional motion along the $x$ axis.
We can recover the parameters of \eqref{modelHam} from (for more details see Appendix \ref{app:wan}):

\begin{equation}
   \label{eq:secondQ}
   \begin{split}
   H=& \int\psi^\dagger(r)\left[- \frac{\hbar^2 \mathbf{\nabla}^2}{2 m} + V_{\textrm{opt}}(\mathbf{r}) \right]\psi(r)+ \\ 
     & +\int \psi^\dagger(\mathbf{r})\psi^\dagger(\mathbf{r}')V(\mathbf{r}'-\mathbf{r})\psi(\mathbf{r}')\psi(\mathbf{r})\textrm{d}^3\mathbf{r}\textrm{d}^3\mathbf{r}'. 
   \end{split}
\end{equation}
The function $V(\mathbf{r})$ represents the sum of contact ($V_c$) and dipolar ($V_d$) interactions, $V(\mathbf{r}) = V_{c}(\mathbf{r}) + V_{d}(\mathbf{r})$, where:

\begin{equation}
  \label{eq:intpots}
  V_c(\mathbf{r}) = \frac{4\pi\hbar^2 a_s}{m} \delta(\mathbf{r}),  \quad V_d(\mathbf{r}) = \frac{C_{dd}}{4\pi} \frac{1 - 3\cos^2 \theta}{r^3},
\end{equation}
with $\theta$ being the angle between the direction of polarization and $\mathbf{r}$, and $a_s$ being the scattering length for effective contact interactions \cite{Sinha2007}.

The value of $C_{dd}$ depends on the strength of dipolar interactions, and has the form:
\begin{equation}
C_{dd} = \begin{cases}
         \mu_0 \mu_m^2, & \text{ for magnetic dipole moment $\mu_m$} \\
         \mu_e^2 / \epsilon_0, & \text{ for electric dipole moment $\mu_e$.}
         \end{cases}
\end{equation}

Further in the text we will be using a~representation of the dipolar interaction strength by a~dimensionless quantity:

\begin{equation}
  \label{eq:dPar}
  d = \frac{m C_{dd}}{2 \pi^3 \hbar^2 a}.
\end{equation}
\begin{figure}
  \includegraphics[width=\columnwidth,keepaspectratio]{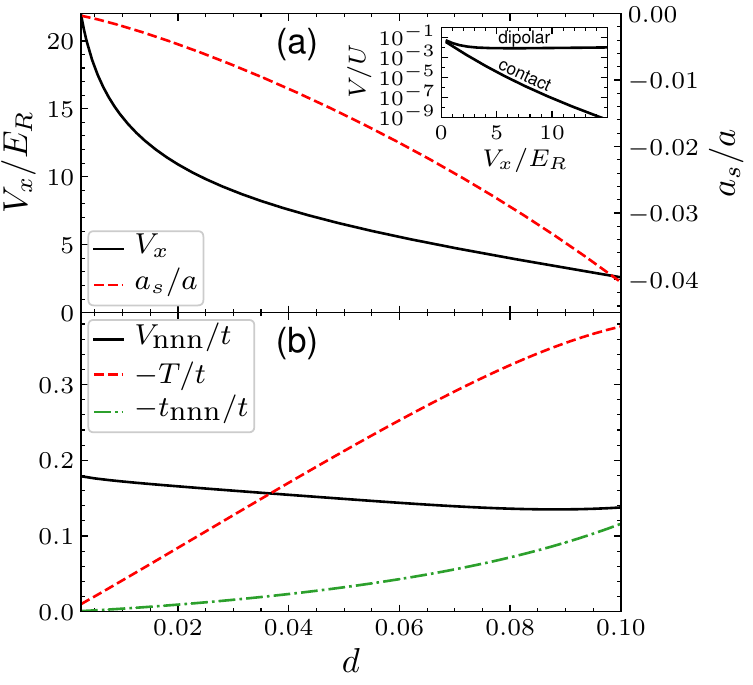}
  \caption{\label{vas_fig} Panel (a) shows values of $V_x$ and $a_s/a$ necessary to get $U/t = 2$ and $V/t = 1.5$ for different values of $d$. Panel (b) shows values of parameters in Hamiltonian \eqref{modelHam} in such case. Inset shows the values of $V/U$ for dipolar-only and contact-only terms respectively.}
\end{figure}
In effect, we have two parameters, $V_x$ and $a_s$ that can be controlled in the  experiment (using previously mentioned Feshbach resonance), and $d$ which depends on the kind of particles used in an experiment (we can, however, modify the strength of dipolar interactions by changing the direction of polarization). In case of molecules, the $d$ can be controlled by the external electric field inducing the dipole moment.
In this work, we set the dipole direction to be perpendicular to that of the lattice, so that dipolar interactions are maximally repulsive.
Then, for a~given values of $U/t$ and $V/t$, the appropriate values of $V_x$ and $a_s$ can be found, which in turn determines the values of $t_{\textrm{nnn}}/t$, $V_{\textrm{nnn}}/t$ and $T/t$.

Let us remark that  one can, in principle, employ a transverse harmonic confinement of the boson gas \cite{Sowinski2012} to change the relative values of the parameters of dipolar interactions.
We have found that while it does provide more control over the values of $T/t$, ultimately they are of a similar magnitude compared to what we obtain solely with $V_{\textrm{opt}}$ and so we refrain from including that method in our considerations.

We denote the values of $V$ and $U$ restricted to contact (dipolar) interactions only as $V_c$ ($V_d$) and $U_c$ ($U_d$).
In the most common parameter range used in this paper, $V/U$ is of the order of 1.
For the optical lattice that we consider (Appendix \ref{app:wan}), both $V_c/U_c$ and $V_d/U_d$ are smaller than $10^{-1}$ (see inset in Fig. \ref{vas_fig}).
Consequently, for a given positive value of $d$, the value of $a_s$ has to be negative in order to lower the value of $U$ to achieve the desired $V/U$.

We now take a closer look into how changes in the dipolar interaction strength influence the validity of using \eqref{modelHam} for a fixed phase diagram point ($U/t$, $V/t$).
$V_d$ and $U_d$ increase linearly with $d$, and so must $|a_s|$ if we want to maintain the desired ratio of $V/U$.
To keep $V/t$ (which is approximately $V_d/t$) and $U/t$ unchanged, the lattice must be made shallower (as $t$ depends solely on $V_x$).
Since the tight-binding approximation is no longer correct for shallow lattices, this provides an effective upper limit for $t$, which gets more strict as $d$ increases.
The maximum value of $d$ we consider in this paper is 0.1 - which corresponds to $V_x$ roughly equal to $2.5 E_R$ for the exemplary values of $U/t = 2$ and $V/t = 1.5$ (see Fig. \ref{vas_fig}, where we also plot the resulting values of $V_{\textrm{nnn}}/t$, $T/t$ and $t_{\textrm{nnn}}/t$).

To give an example of the magnitude of $d$ for real atoms and molecules, we first assume the lattice constant to be $a = 532 \mathrm{nm}$.
Single atoms have weak dipole moments (for $^{52}\textrm{Cr}$: $d \approx 9.7 \times 10^{-4}$, for $^{168}\textrm{Er}$: $d \approx 4.3 \times 10^{-3}$ and for $^{164}\textrm{Dy}$: $d \approx 8.5 \times 10^{-3}$)\cite{Lewenstein2012,Aikawa2012,Lu2011}.
The values for molecules can be a~few orders of magnitude greater (for $^{168}\textrm{Er}_2$: $d \approx 0.1$)\cite{Frisch2015}.
It is worth noting that multiple different experimental methods of decreasing $a$ in optical lattices [which would increase $d$, see Eq. \eqref{eq:dPar}] by a factor of 2 or 3 (with a prospect for a larger value) have been developed and tested \cite{Salger2007,Yi2008,Sun2011,Nascimbene2015}.

\section{The phase transitions at $\rho = 1$}\label{sec:rh1}

The full phase diagram calculated numerically for the EBH model with $t$, $U$ and $V$ as the only parameters and a unit mean density $\rho = 1$ has been studied in detail already~\cite{Rossini2012,Batrouni2014} and here we will only briefly sum up the possible phases observed in the ($V/t$, $U/t$) plane.
For large values of $t$, the system is in the SF phase, whereas large values of $U/t$ with small $V/t$ drive the system into the MI.   Large enough values of $V/t$ for a sufficient $U/t$ put the system in the DW phase.
The HI is present on the phase diagram in between the three previously mentioned phases, that is for intermediate values of both $V/t$ and $U/t$.

In this section we will calculate how the locations of the transitions between these phases change for the Hamiltonian \eqref{modelHam}, depending on dipolar interaction strength $d$.
We will not, however, recover a full phase diagram, but instead we focus on two lines, given by the constraints $V/U = 0.75$ and $U/t = 3$.
The first of these values is chosen because it covers three of the phases achievable in EBH model (DW, HI and SF) and has been already extensively analyzed\cite{Batrouni2014,Gremaud16}, while the second one allows us to examine the MI phase (in addition to DW and HI, which are also present in that case).

In order to determine the boundaries between different phases, we define their characteristic properties:
\begin{enumerate}
  \item DW: $\mathcal{O}_{DW} \neq 0$, $\Delta E \neq 0$,
  \item MI: $\mathcal{O}_{DW} = 0$, $\mathcal{O}_{\textrm{string}} = 0$, $\Delta E \neq 0$,
  \item HI: $\mathcal{O}_{DW} = 0$, $\mathcal{O}_{\textrm{string}} \neq 0$, $\Delta E \neq 0$,
  \item SF: $\mathcal{O}_{DW} = \mathcal{O}_{\textrm{string}} = 0$, $\Delta E = 0$,
\end{enumerate}
with order parameters defined similarly as in~\cite{Rossini2012}: $\mathcal{O}_{p} \equiv \lim_{r\to\infty} C_{p}$, for the following correlators:
\begin{align}
  \label{eq:csf} C_{SF}(r)  = {}& \braket{b_j^{\dag} b_{j+r}} \\
  \label{eq:cdw} C_{DW}(r)  = {}& (-1)^r \braket{\delta n_j \delta n_{j+r}} \\
  \label{eq:cst} C_{\textrm{string}}(r)  = {}& \braket{\delta n_j e^{i \pi \sum_{j\leq k \leq j+r} \delta n_k} \delta n_{j+r}},
\end{align}
where $\delta n_j = n_j - \rho$.
The energy gap and its thermodynamic limit extrapolation are defined simply as $\Delta E(L) = E^{(1)}(L) - E^{(0)}(L)$ and $\Delta E = \lim_{L\to\infty} \Delta E(L)$, where $E^{(k)}(L)$ is energy of $k-$th excited state in the lattice of length $L$ ($k=0$ is the ground state).

We will be also using the fact that for the superfluid phase it can be shown, using the Luttinger liquid theory, that the correlations in the system show power-law decay \cite{giamarchi2004}:
\begin{equation}
  \label{SFcor}
  C_{SF}(r) \sim r^{-K/2}.
\end{equation}

\subsection{$V/U = 0.75$ constraint} \label{ssec:vu34}
\begin{figure}
  \includegraphics[width=\columnwidth,keepaspectratio]{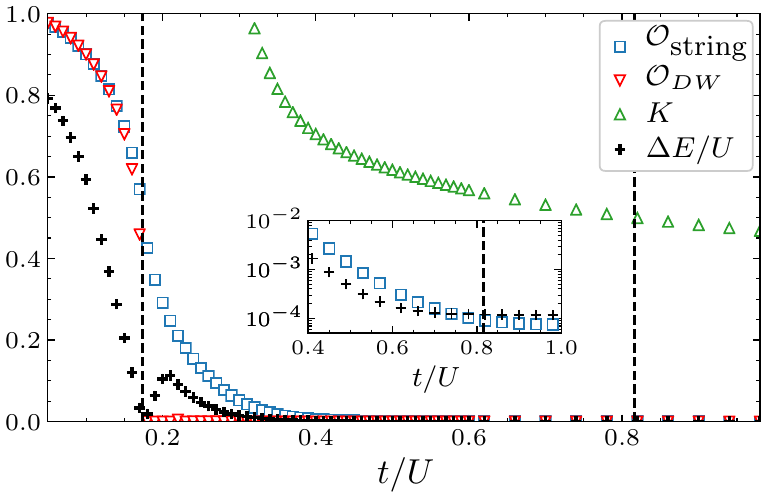}
  \caption{\label{rho1_vu34_ops_fig} (color online) The values of the string and DW order parameters, critical exponent $K$ and energy gap $\Delta E$ for $V/U = 3/4$, $d = 0.02$. The positions of black, dashed, vertical lines correspond to the critical values of $t/U$ for DW-HI and HI-SF transitions ($t_c^{DW-HI}/U \approx 0.175$ and $t_c^{HI-SF} \approx 0.82$). The inset shows a logarithmic plot of $\mathcal{O}_{\textrm{string}}$ and $\Delta E$ near HI-SF transition.}
\end{figure}

We present the results of our calculations for the model \eqref{modelHam} obtained using the DMRG method described in Appendix \ref{app:dmrg}.
For $t/U$ close to zero the system is in the DW phase.
As the value of $t/U$ is increased, the first transition is a DW-HI transition at $t_c^{DW-HI}/U$.
The transition location can be easily determined, as for $t=t_c^{DW-HI}$: 1) the gap $\Delta E$ closes and 2) the order parameter $\mathcal{O}_{DW}$ vanishes (see Fig.~\ref{rho1_vu34_ops_fig}, where the values of the order parameters are plotted for $d=0.02$).
$\Delta E$ is linear with respect to $t/U$ at both sides of the transition which allows us to easily determine where the gap closes.
Additionally, a~function $a [(t-t_c)/U]^{-b}$ can be fitted to the numerically computed $\mathcal{O}_{DW}$ near the transition point for $t/U < t_c^{DW-HI}/U$.
The values of $t_c/U$ obtained with these methods are in agreement with each other (with a difference of less than $5\times 10^{-3}$ for every value of $d$ that was considered).

For even larger $t$, the consecutive transition occurs between the HI and the SF phases, but the determination of its location, $t_c^{HI-SF}/U$, proves to be more difficult.
As in the earlier case, the energy gap closes and the appropriate order parameter ($\mathcal{O}_{\textrm{string}}$) goes to zero.
However, the decay of both $\Delta E$ and $\mathcal{O}_{\textrm{string}}$ features an exponential tail and does not provide a~clear value of the transition point (see inset of Fig. \ref{rho1_vu34_ops_fig}).
In order to determine the correct value, we fit the correlations $C_{SF}(r)$ for each $L$ according to \eqref{SFcor} and then extrapolate the obtained $K$ to $L\to\infty$ limit.
It has been shown \cite{Kuhner2000}, that $K = 0.5$ for $\rho=1$ at the transition between insulator and superfluid phases.  That is the criterion we use here to determine $t_c^{HI-SF}/U$.

\begin{figure}
  \includegraphics[width=\columnwidth,keepaspectratio]{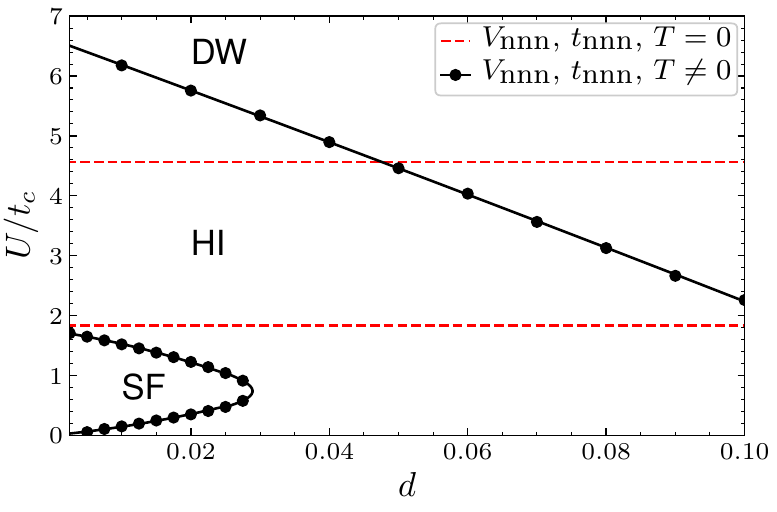}
  \caption{\label{rho1_vu34_fig} (color online) Critical values of $U/t$ for DW-HI and HI-SF transitions, $V/U = 3/4$ (black, solid lines), same for model with $V_{\textrm{nnn}}$, $t_{\textrm{nnn}}$ and $T$ set to 0 (red, dashed lines).}
\end{figure}

The results of the analysis described above are shown in Fig. \ref{rho1_vu34_fig}, where the dependence on the chosen $d$ value for both DW-HI and HI-SF transitions is plotted as black solid lines.
The results of similar calculations, but with parameters $V_{\textrm{nnn}}$, $t_{\textrm{nnn}}$ and $T$ set to 0 are marked with the vertical red dashed lines. 
$U/t_c$ value for DW-HI transition has a~strong, linear dependence on $d$, and the transition point is moved considerably both for small and large values of $d$ in the chosen interval ($0 < d \le 0.1$).
The situation is different for the HI-SF transition - while for values of $d$ close to 0, $U/t_c$ is almost the same as for an ordinary EBH, the SF phase disappears completely around $d = 0.03$.
What can also be seen for the intermediate values of $d$ is that for small $U/t_c$ an another transition appears - in simulations we see the re-emergence of HI phase, indicated by a~rise of $\mathcal{O}_{\textrm{string}}$, $\Delta E$ and $K$ (the transition point is once again pinpointed by equation $K = 0.5$). The striking substantial difference between the two models indicate that a real care has to be taken when applying the tight-binding approximate Hamiltonian to a given physical system.

\subsection{$U/t = 3$ constraint} \label{ssec:ut3}

\begin{figure}
  \includegraphics[width=\columnwidth,keepaspectratio]{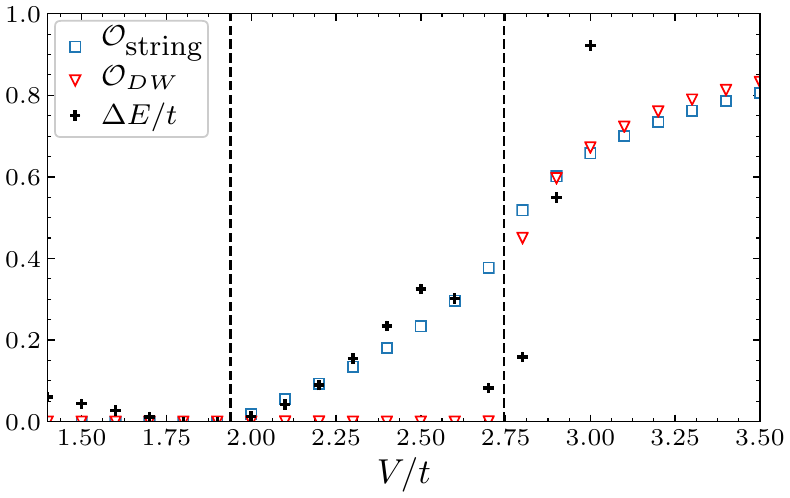}
  \caption{\label{rho1_ut3_ops_fig} (color online) The values of the order parameters \eqref{eq:csf}, \eqref{eq:cdw}, \eqref{eq:cst} for $U/t = 3$, $d = 0.09$. The positions of black, dashed, vertical lines correspond to the critical values of $V/t$ for DW-HI and HI-MI transitions ($V/t_c^{HI-MI} \approx 1.94$ and $V/t_c^{DW-HI} \approx 2.74$).}
\end{figure}

\begin{figure}
  \includegraphics[width=\columnwidth,keepaspectratio]{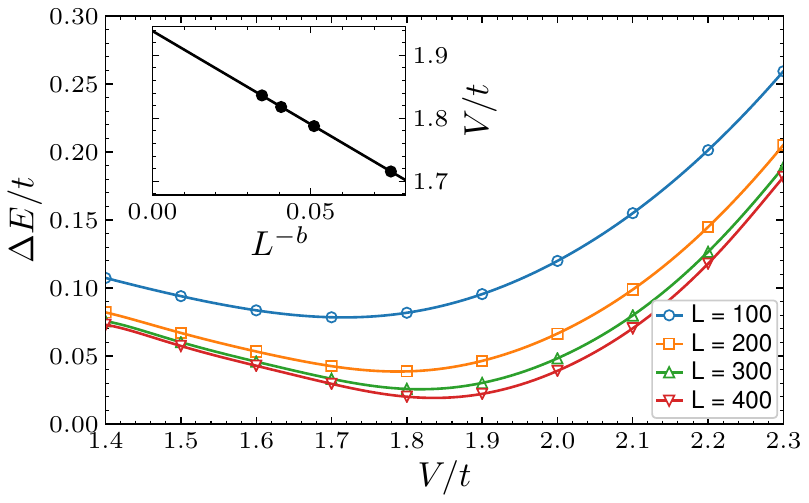}
  \caption{\label{rho1_ut3_egap_fig} (color online)  The energy gap for different system sizes and $U/t = 3$, $d = 0.09$. Inset shows extrapolation for $L\to\infty$, $b \approx 0.56151$.}
\end{figure}

In this case, there exist two transitions between three insulating phases: DW-HI and HI-MI.
The method of locating HI-DW transition is the same as in \ref{ssec:vu34} (corresponding plot of order parameters for $U/t = 3$ and $d=0.09$ is shown in Fig. \ref{rho1_ut3_ops_fig}).
For the HI-MI transition a different approach must be undertaken, as $\Delta E$ does not have a~linear dependence on $t$ near the transition point.
To determine $V/t_c$ we find the minimum of $\Delta E$ with respect to $V/t$ for each available $L$, and then we extrapolate it for $L\to\infty$ using a~power function $a L^{-b} + V/t_c$ (see Fig.~\ref{rho1_ut3_egap_fig}).

\begin{figure}
  \includegraphics[width=\columnwidth,keepaspectratio]{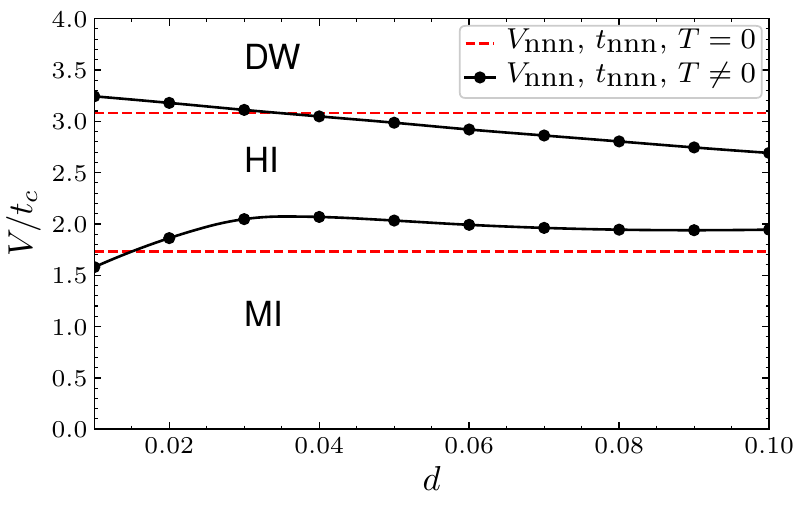}
  \caption{\label{rho1_ut3_vu34_fig} (color online)  Critical values of $V/t$ for DW-HI and HI-SF transitions, $U/t = 3$ (black, solid lines), same for model with $V_{\textrm{nnn}}$, $t_{\textrm{nnn}}$ and $T$ set to 0 (red, dashed lines).}
\end{figure}

We plot the results in Fig.~\ref{rho1_ut3_vu34_fig} comparing them with the results obtained for a pure EBH model i.e. setting $V_{\textrm{nnn}}$, $t_{\textrm{nnn}}$ and $T$ in \eqref{modelHam} to 0.
While the changes are not as drastic as for fixed $V/U = 0.75$, the HI phase gets narrower with respect to $V/t$ as $d$ increases.

\section{The phase diagram for $d=0.1$}\label{sec:rhv}

\begin{figure}
  \includegraphics[width=\columnwidth,keepaspectratio]{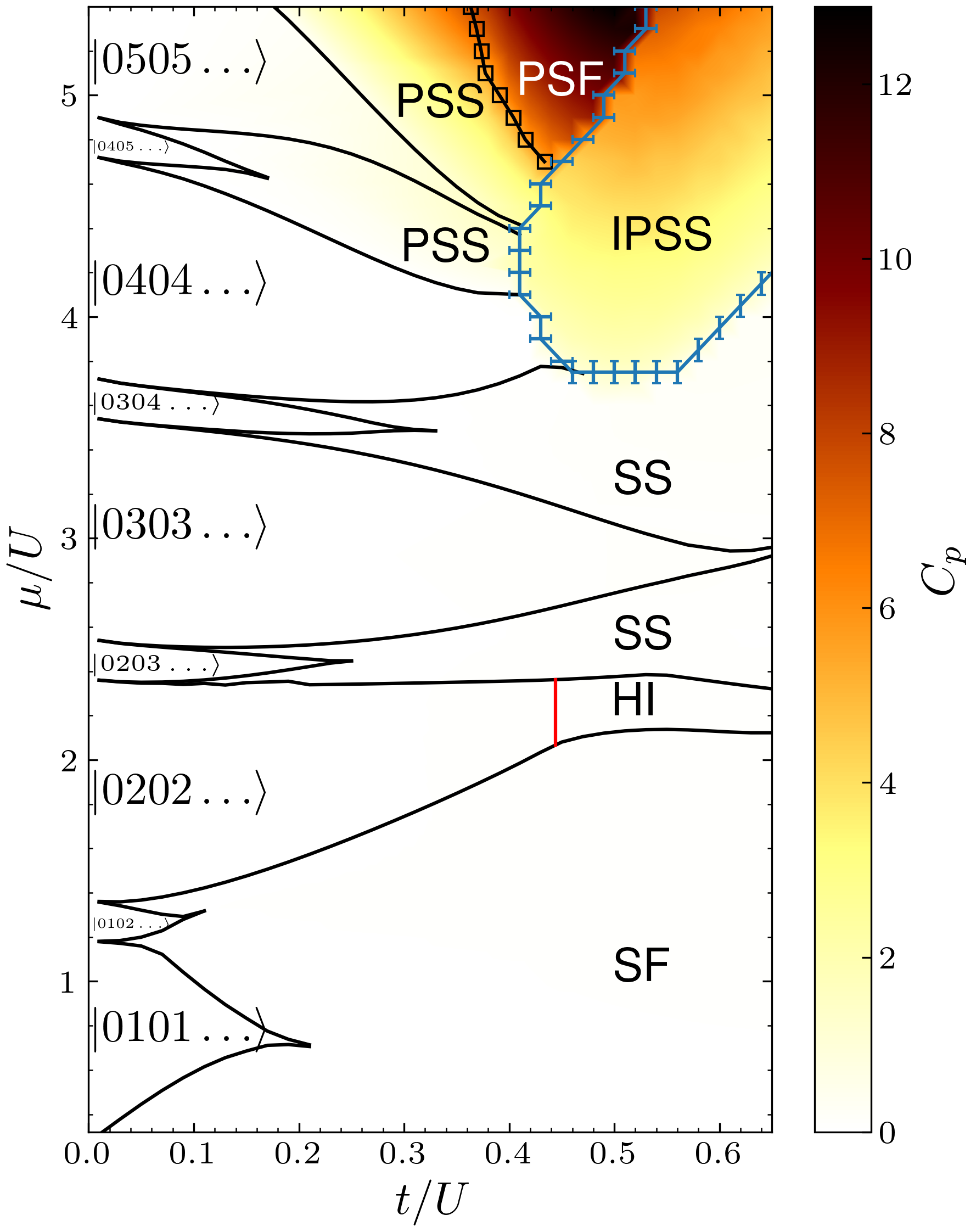}
  \caption{\label{brhoPCor_fig} (color online)  The phases for the system for $d = 0.1$ at a~fixed ratio $V/U = 0.75$. Black lines showing the boundaries of DW phases are the values of $\mu_+$ and $\mu_-$ obtained from OBC DMRG ($L=200$). The black squares come from sine-square deformation (SSD) DMRG (see Appendix \ref{app:ssd} for details) for $L=100$  and show the transition points between PSS and PSF (where $\mathcal{O}_{DW}$ vanishes). Blue error bars mark the boundaries of the IPSS phase (also SSD DMRG, $L=100$). The value of pair-tunneling correlations $C_p$ \eqref{eq:cp} is plotted as a color map with a scale shown on the right.}
\end{figure}

In this section, we will characterize the phase diagram without constraining the density of particles $\rho$ while setting $V/U = 0.75$ and $d = 0.1$.
The results for an ordinary EBH model, obtained mostly using quantum Monte Carlo methods can be found in \cite{Batrouni2014,Kawaki17}.	
To this end we calculate the ground state energies using DMRG with OBC (for technical details see Appendix \ref{app:dmrg}) for $\rho$ corresponding to each of the DW phases present in the system for vanishing tunnelings. It is easy to convince oneself that the DW phase requires a commensurate relation between the number of particles and number of sites. Restricting to next-nearest neighbor interactions the corresponding densities are $\rho_{DW} = n_{DW}/4$, where $n_{DW}\geq2$, $n_{DW}\in \mathbb{Z}$.
Repeating the same calculations with particles added/removed from the system allows us to obtain the chemical potential: $\mu(N, L) = \partial E(N, L)/{\partial N}$.
We can then get the boundaries of DW phases, as a~discontinuity in $\mu(N, L)$ at $N_{DW}=\rho_{DW} L$.
The lower boundary for the DW phase is then given by $\mu^- = \lim_{N\to N_{DW}^+} \mu(N, L)$, while the upper one by $\mu^+ = \lim_{N\to N_{DW}^-} \mu(N, L)$.
By adjusting the system size we have verified that systems with $L=200$ are sufficiently large to properly determine the values of $\mu_L$ and $\mu_U$, for most of the boundary $\mu^- = E(N, L) - E(N-1, L)$, $\mu^+ = E(N+1, L) - E(N, L)$ (the only exception are the cusps at the rightmost edges of DW lobes, where we take into account $E(N-2, L)$ and $E(N+2, L)$ and do the quadratic interpolation).
The resulting phase diagram can be seen in Fig. \ref{brhoPCor_fig}.
We remark that apart from the conventional $\ket{0(2\rho)0(2\rho)0\dots}$ DW phases, with $\rho=\rho_{DW}$ we observe $\ket{0\left(2\rho-\frac{1}{2}\right)0\left(2\rho+\frac{1}{2}\right)\dots}$ phases for odd $n_{DW}$ as an effect of introducing $V_{\textrm{nnn}}n_i n_{i+2}$ coupling terms to the Hamiltonian. The corresponding DW regions are, fortunately, quite tiny, showing that for most parameters, the picture obtained within EBH model is correct.

\begin{figure}
  \includegraphics[width=\columnwidth,keepaspectratio]{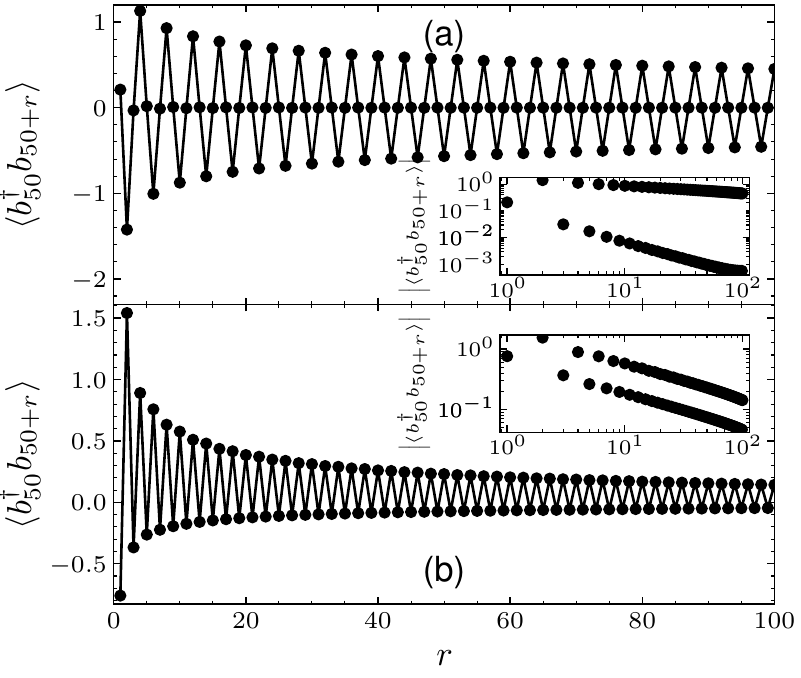}
  \caption{\label{sscor_fig} OBC DMRG results of $\braket{b^{\dag}_i b_{i+r}}$ correlations in the middle of $L=200$ lattice  at (a) $\rho = 1.25$, $t/U = 0.59$ (SS phase) and (b) $\rho = 2.25$, $t/U = 0.37$ (PSS phase). Log-log plots of the same correlations are shown in the insets.}
\end{figure}

Apart from the abundant DW phases we observe either SF- or SS-like phases as indicated by the power law decay of $C_{SF}$ correlations \eqref{SFcor}.
The difference between the two phases is a~nonzero density-wave order parameter value in the supersolid phase.
The trivial SF phase is seen for $\rho < 1$, however, we observe the emergence of pair superfluid (PSF) phase for large enough $\mu$.
We use the pair-tunnelling correlation:
\begin{equation}
  C_p = \frac{1}{L}\sum_{i}\braket{b_i^{\dag} b_i^{\dag} b_{i+1} b_{i+1}}
  \label{eq:cp}
\end{equation}
as a~measure of pair-superfluidity [see Fig. \ref{brhoPCor_fig}].
The phases marked as SS and PSS (pair-supersolid) in Fig. \ref{brhoPCor_fig} differ from conventional supersolid phases in a simple EBH model, where $C_{SF}(r)$ is always positive.
$C_{SF}(r)$ is negative for $r = 4 n + 2$, $n\in\mathbb{Z}$ in SS phase [Fig. \hyperref[sscor_fig]{\ref*{sscor_fig}(a)}] and for odd $r$ in the PSS phase [Fig. \hyperref[sscor_fig]{\ref*{sscor_fig}(b)}].
The other difference is that $C_p > 0$ in the PSS phase.
We remark that both PSS and PSF phases have been previously observed in numerical calculations for EBH Hamiltonians with density-dependent tunneling \cite{Sowinski2012,Jiang2012,Rapp2012}.

Next, we are going to describe the last phase present in the phase diagram, which we will call an incommensurate pair supersolid (IPSS).
This phase is characterized by a finite, positive $C_p$ and a structure factor:
\begin{equation}
  \label{eq:sfac}
  S(q) = \frac{1}{L^2}\sum_{j, k = 1}^L \braket{n_j n_k} e^{-i q (j - k)}
\end{equation}
having a~peak at $\pi/2 < q < \pi$, which is incommensurate with respect to lattice size and the particle density.
In order to identify this phase, we use the sine-squared deformation (SSD) variant of the DMRG method which we describe in Appendix \ref{app:ssd}.

\begin{figure}
  \includegraphics[width=\columnwidth,keepaspectratio]{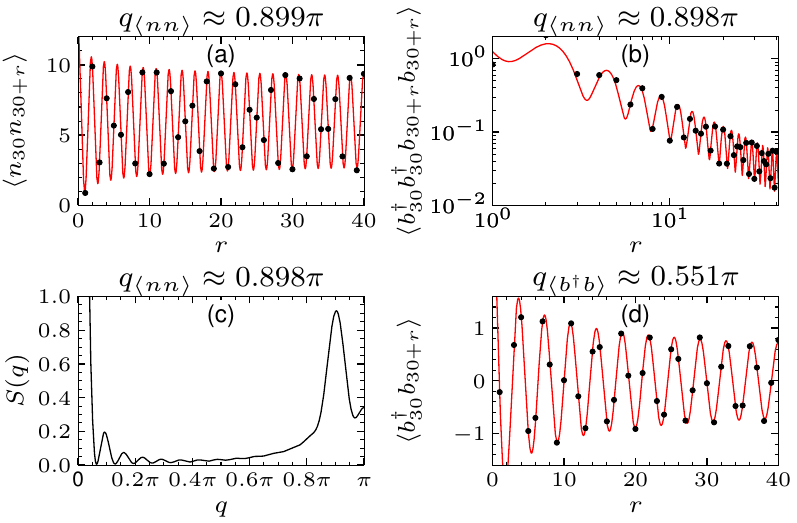}
  \caption{\label{IPSS_fig} (color online) Correlations and structure factor values obtained by SSD DMRG for the system in IPSS phase ($L = 100$, $t/U = 0.48$ and $\mu = 3.7$). (a) Density correlations, (b) pair correlations, (c) structure factor \eqref{eq:sfac} and (d) creation-annihilation correlations. For (a), (b) and (d), black points mark the numerical results with red lines showing the fits of the functions in \cref{eq:density_cor,eq:pair_cor,bdagb_cor}. The value of the appropriate wavenumber $q_{\alpha}$ obtained from the fits [or from the position of the $S(q)$ peak in (c)] is written above each of the subplots.}
\end{figure}

In IPSS we see periodic modulation of both density and density-density correlations [Fig. \hyperref[IPSS_fig]{\ref*{IPSS_fig}(a)}] in the form of:
\begin{align}
  \braket{n_i} & = \rho_{\textrm{bulk}} + \Delta \rho \sin (q_{\braket{nn}} i + \varphi_0), \label{eq:density} \\
  \braket{n_i n_{i+r}} & = C_1 + A_1\sin (q_{\braket{nn}} r + \varphi_1) r^{-\alpha_1}, \label{eq:density_cor}
\end{align}
where $q_{\braket{nn}}$ is the same wavenumber value for which there is a~peak in $S(q)$ [see Fig. \hyperref[IPSS_fig]{\ref*{IPSS_fig}(c)}]
The pair correlations are also showing the same modulation, while at the same time following a power-law decay [Fig. \hyperref[IPSS_fig]{\ref*{IPSS_fig}(b)}]:
\begin{equation}
  \label{eq:pair_cor}
  \braket{b_i^{\dag} b_i^{\dag} b_{i+r} b_{i+r}} = [C_2 + A_2 \sin(q_{\braket{nn}} r + \varphi_2)] r^{-\alpha_2}.
\end{equation}

Another modulation can be observed in $\braket{b_i^{\dag} b_{i+r}}$, however in this case the wavenumber differs from $q_{\braket{nn}}$ and the values oscillate around 0 [See Fig. \hyperref[IPSS_fig]{\ref*{IPSS_fig}(d)}]:
\begin{equation}
  \label{bdagb_cor}
  \braket{b_i^{\dag} b_{i+r}} = A_3 \sin(q_{\braket{b^{\dag}b}} r + \varphi_3) r^{-\alpha_3}.
\end{equation}

\begin{figure}
  \includegraphics[width=\columnwidth,keepaspectratio]{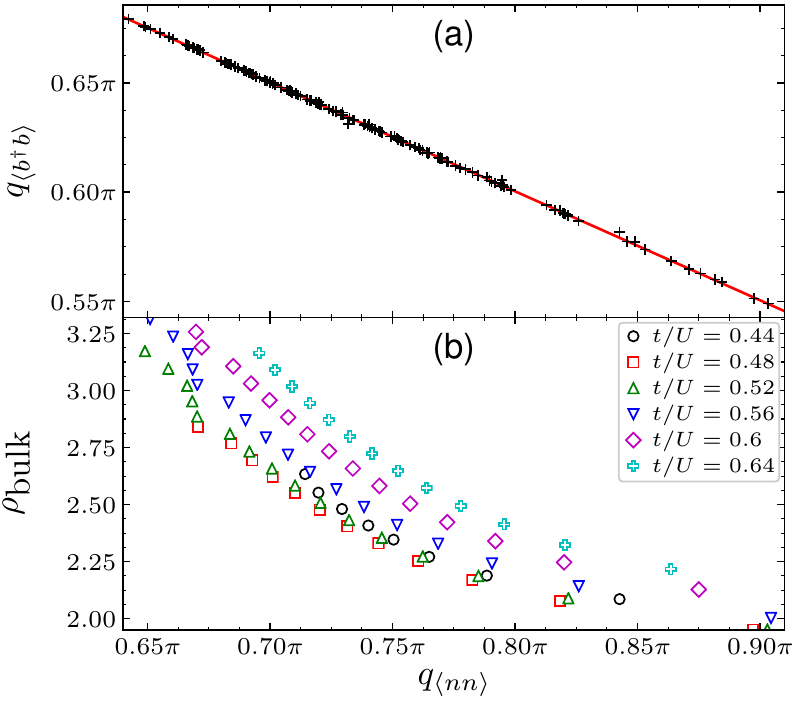}
  \caption{\label{kkrho_fig} (color online) The results of SSD DMRG for the IPSS phase. (a) The relation between $q_{\braket{nn}}$ and $q_{\braket{b^{\dag}b}}$. Linear fit (red) is $q_{\braket{b^{\dag}b}} = 0.9991(6)\pi - 0.4984(7) q_{\braket{nn}}$. (b) The relation between $q_{\braket{nn}}$ and $\rho_{\textrm{bulk}}$ shown for different values for $t/U$.}
\end{figure}

After combining the results for many different $\mu$ and $t/U$ parameters, we can provide the relation between $q_{\braket{b^{\dag}b}}$ and $q_{\braket{nn}}$: [see Fig. \hyperref[kkrho_fig]{\ref*{kkrho_fig}(a)}]
\begin{equation}
  \label{qbb_qnn}
  q_{\braket{b^{\dag}b}} = \pi - 0.5 q_{\braket{nn}}.
\end{equation}

We also note that is $q_{\braket{nn}}$ does not depend exclusively on $\rho_{bulk}$ [see Fig. \hyperref[kkrho_fig]{\ref*{kkrho_fig}(b)}](which is the case in e.g. underdoped $\rho=0.5$ DW, where $q = 2\pi\rho$ \cite{Gremaud16})

\section{Conclusions}\label{sec:con}
In this paper we have presented the accurate Hamiltonian representation of a~one dimensional system of bosons in an optical lattice considering both the dipolar and the contact interactions (the mutual strength of which may be balanced using the Feshbach resonance).
We have employed the well established DMRG method to measure the dependence of the phase transitions on often overlooked terms in the EBH model (most notably the next-nearest neighbor tunnelings  and the density-dependent tunnelings).
We observe the suppression of the SF phase with rising dipolar interaction strength.
In the case of fixed $\rho = 1$ we also note a~stable presence of a~nontrivial, highly non-locally correlated HI phase throughout the considered parameters range, which is even more pronounced for a~realistic, low values of dipolar interactions. This robustness can be traced back to the fact that HI is a symmetry protected topological state \cite{Wierschem14}. 

For a greater dipolar interaction strength and higher densities we observe interesting pair-correlated phases.
Among those, we put a~particular emphasis on characterizing a novel incommensurate pair superfliud phase, whose distinctive feature is an incommensurate density wave order.
That phase is not present in neither standard EBH model nor for large dipole-dipole interactions in small diagonalization studies.
We also notice a~particular relation between wavenumbers characterizing different correlations measured in this phase \eqref{qbb_qnn} which may provide some insight into how to construct an appropriate theoretical description.
Rigorous theoretical treatment of IPSS is, however, beyond the scope of this paper.

\acknowledgments

This work was realized under National Science Center (Poland) projects No. 2015/19/B/ST2/01028 (KB), 2016/23/D/ST2/00721 (M\L) and  2016/21/B/ST2/01086 (JZ). It  was also supported in part by PL-Grid Infrastructure as well as by the EU Horizon2020 FET project QUIC (nr. 641122).

\appendix

\section{The determination of Hamiltonian parameters}\label{app:wan}
The values of the parameters in the model \eqref{modelHam} have been calculated numerically using Wannier function representation for periodic boundaries system with a~standard optical lattice potential $V_{\textrm{opt}}(\mathbf{r})$.
In a~numerical calculations described below we assume the lattice is in the form of a~cube with $N^3$ sites, so that the total volume $\Omega = (N a)^3$, where $a = \pi / k_L$ is the lattice constant.

The Bloch functions of the form:
\begin{equation}
  \label{eq:bloch}
  \phi_{\mathbf{k}}(\mathbf{r}) = e^{i \mathbf{k} \cdot \mathbf{r}} u_k(\mathbf{r}),
\end{equation}
where $u_k(\mathbf{r})$ is a function with the same periodicity as the lattice potential, are calculated for non-interacting Hamiltonian, $H_{\textrm{NI}} = \frac{-\hbar^2 \mathbf{\nabla}^2}{2 m} + V_{\textrm{opt}}(\mathbf{r})$, as the lowest energy eigenvectors of the Shr\"odinger equation:
\begin{equation}
  \label{eq:blochShr}
  H_{\textrm{NI}} \phi_{\mathbf{k}}(\mathbf{r}) = E_k \phi_{\mathbf{k}}(\mathbf{r}).
\end{equation}

Wannier functions can be calculated in a~usual way~\cite{Kohn1959} from the Bloch functions: 

\begin{equation}
  \label{eq:wannier}
  w_{n}(\mathbf{r}) = \frac{1}{\sqrt{N^3}}\sum_{\mathbf{k}\in\textrm{BZ}} \phi_{\mathbf{k}}(\mathbf{r}) e^{-i k_x a n},
\end{equation}
where $\phi_{\mathbf{k}}(0)$ is real and positive and $n$ is the number of a lattice site in $x$ direction (we assume $y = z = 0$) and the summation is done over $\mathbf{k} = (k_x, k_y, k_z)$ from the first Brillouin zone.

Substituting the field operators of the form $\phi(\mathbf{r}) = \sum_i w_i(\mathbf{r}) b_i$ to \eqref{eq:secondQ} we get:
\begin{align}
  \label{eq:HamPars1}
  t        = {}& t_{i(i+1)} \nonumber \\
  t_{\textrm{nnn}} = {}& t_{i(i+2)} \nonumber \\
  U        = {}& V_{iiii} \nonumber \\
  V        = {}& V_{i(i+1)i(i+1)} + V_{i(i+1)(i+1)i} \nonumber \\
  V_{\textrm{nnn}}  = {}& V_{i(i+2)i(i+2)} + V_{i(i+2)(i+2)i} \nonumber \\
  T        = {}& -0.5[V_{ii(i+1)i} + V_{iii(i+1)}],
\end{align}
with:

\begin{align}
  \label{eq:HamPars2t}
  t_{ij}   = {}& -\int_{\Omega} d \mathbf{r} w_i^*(\mathbf{r}) H_{\textrm{NI}}\, w_{j}(\mathbf{r}) \\
  \label{eq:HamPars2V}
  \begin{split}
    V_{ijkl} = & \int_{\Omega} d\mathbf{r_1} d\mathbf{r_2} w_i^*(\mathbf{r_1}) w_j^*(\mathbf{r_2}) \\
               & \times V (\mathbf{r_1} - \mathbf{r_2}) w_k(\mathbf{r_1}) w_l(\mathbf{r_2}).
  \end{split}
\end{align}

Integral \eqref{eq:HamPars2t} is straightforward to calculate using \eqref{eq:blochShr} and \eqref{eq:wannier}.
In order to calculate \eqref{eq:HamPars2V}, we use periodic extension of the interaction potential:

\begin{equation}
  V(\mathbf{r}) = \frac{1}{\Omega}\sum_{\mathbf{k}} \widetilde{V}(\mathbf{k}) e^{i \mathbf{k} \cdot \mathbf{r}},
\end{equation}
where $\mathbf{k} = \frac{2\pi}{N a} (n_1, n_2, n_3)$, $n_i \in \mathbb{N}$ and $\widetilde{V}(\mathbf{k}) = \widetilde{V_c}(\mathbf{k}) + \widetilde{V_d}(\mathbf{k})$ is the sum of the Fourier transforms of the contact and dipolar interaction potentials \eqref{eq:intpots}:

\begin{equation}
  \label{eq:intpotsF}
  \widetilde{V_c}(\mathbf{k}) = \frac{4\pi\hbar^2 a_s}{m},  \quad \widetilde{V_d}(\mathbf{k}) = C_{dd} (\cos^2 \gamma - 1/3),
\end{equation}

where $\gamma$ is the angle between the direction of polarization and $\mathbf{k}$.
For convenience, we group the Wannier functions with the same arguments $w_{ij}(\mathbf{r}) = w_i^*(\mathbf{r}) w_j (\mathbf{r})$:
\begin{equation}
  \label{eq:Vbeg}
  \begin{split}
    V_{ijkl} = & \int_{\Omega}d\mathbf{r_1} w_{ik}(\mathbf{r_1}) \int_{\Omega} d\mathbf{r_2} V(\mathbf{r_1} - \mathbf{r_2}) w_{jl}(\mathbf{r_2}) \\
             = & \int_{\Omega}d\mathbf{r_1} w_{ik}(\mathbf{r_1}) (V * w_{jl})(\mathbf{r_1}) \\
             = & \frac{1}{\Omega} \int_{\Omega}d\mathbf{r_1} w_{ik}(\mathbf{r_1}) \sum_{\mathbf{k_2}} \widetilde{(V*w_{jl})}(\mathbf{k_2}) e^{i \mathbf{k_2} \cdot \mathbf{r}}.
  \end{split}
\end{equation}
We use the convolution theorem for the Fourier series to obtain:

\begin{equation}
  \label{eq:Vend}
  \begin{split}
    V_{ijkl} = & \frac{1}{\Omega} \int_{\Omega}d\mathbf{r}\, w_{ik}(\mathbf{r}) \sum_{\mathbf{k_2}} \widetilde{V}(\mathbf{k_2}) \widetilde{w_{jl}}(\mathbf{k_2}) e^{i \mathbf{k_2} \cdot \mathbf{r}} \\
             = & \frac{1}{\Omega^2} \int_{\Omega} d \mathbf{r} \sum_{\mathbf{k_1}} \widetilde{w_{ik}}(\mathbf{k_1}) e^{i \mathbf{k_1} \cdot \mathbf{r}} \sum_{\mathbf{k_2}} \widetilde{V}(\mathbf{k_2}) \widetilde{w_{jl}}(\mathbf{k_2}) e^{i \mathbf{k_2} \cdot \mathbf{r}} \\
             = & \frac{1}{\Omega^2} \sum_{\mathbf{k_1}, \mathbf{k_2}} \widetilde{w_{ik}}(\mathbf{k_1}) \widetilde{V}(\mathbf{k_2}) \widetilde{w_{jl}}(\mathbf{k_2}) \int_{\Omega} d\mathbf{r}\, e^{i (\mathbf{k_1} + \mathbf{k_2}) \cdot \mathbf{r}} \\
             = & \frac{1}{\Omega} \sum_{\mathbf{k}} \widetilde{w_{ik}}(-\mathbf{k}) \widetilde{V}(\mathbf{k}) \widetilde{w_{jl}}(\mathbf{k}).
  \end{split}
\end{equation}

\section{DMRG parameters}\label{app:dmrg}
All of the numerical calculations reported in this paper were done using Density Matrix Renormalization Group (DMRG) implementation found in ITensor library~\cite{ITensor}.
For most of the work OBC were used, with sizes from $L = 100$ to $L = 400$ and a~maximum bond dimension $\chi = 600$.
Cutoff $\epsilon$ was set to $10^{-12}$ [$\epsilon$ determines the number of singular values discarded after each singular value decomposition step in ITensor algorithm: $(\sum_{n \in \textrm{discarded}} \lambda_n^2) / \left(\sum_n \lambda_n^2 \right) < \epsilon$].
In Section \ref{sec:rh1}, we limit the maximum number of particles on each lattice site ($N_{\textrm{cut}}$) to~5, while for the OBC and SSD DMRG used in Section \ref{sec:rhv} respectively up to~10 and~12.

Unless stated otherwise, boundary term equal to $2 \rho (n_1 V + n_2 V_{\textrm{nnn}} + n_L V_{\textrm{nnn}})$ was added to break the degeneracy of DW state (the added term simulates a situation, where we have 4 additional sites at the boundaries, with fixed $n_{-1} = 0$, $n_0 = 2\rho$, $n_{L+1} = 0$ and $n_{L+2} = 2\rho$, as expected in one of the DW ground states).
Another motivation for adding these terms is to remove excitations on the edges in the HI phase.

\section{The description of sine-squared deformation DMRG}\label{app:ssd}

\begin{figure}
  \includegraphics[width=\columnwidth,keepaspectratio]{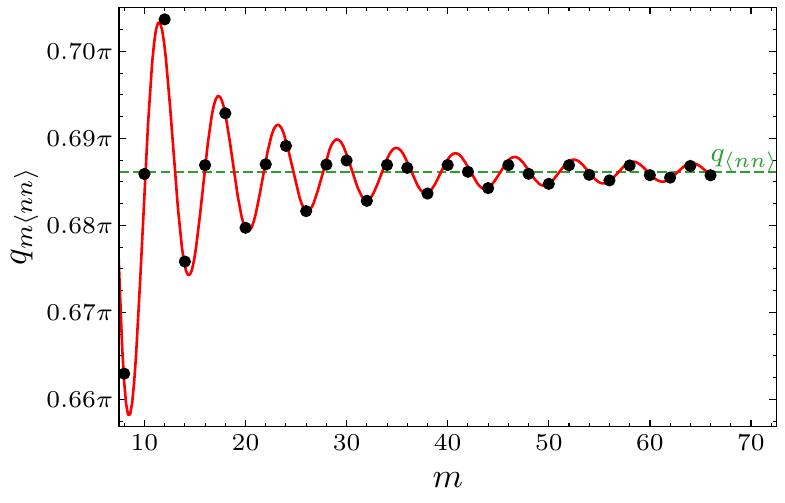}
  \caption{\label{fig:qm} (color online) Position of the peak in $S(q)$ (\ref{eq:sfac}) computed using $m$ middle sites (black points). Red, solid line shows a fit of the form $C + A\, m^{-B} \cos(K m + \phi)$. Here $K\approx 0.342\pi$, which is roughly half of $q_{\braket{nn}} = C \approx 0.686\pi$ (shown as a green, dashed line). At most 2/3 of the all lattice sites have been considered. Data calculated for $\mu/U = 5.2$, $t/U = 0.6$, $L=100$ and $N_{\textrm{cut}} = 12$. The damped oscillations amplitude $S(q)$ position is approximately order of magnitude smaller than FWHM of the $S(q)$ function, which for maximal $m$ is $\approx 0.03\pi $.}
\end{figure}

\begin{figure}
  \includegraphics[width=\columnwidth,keepaspectratio]{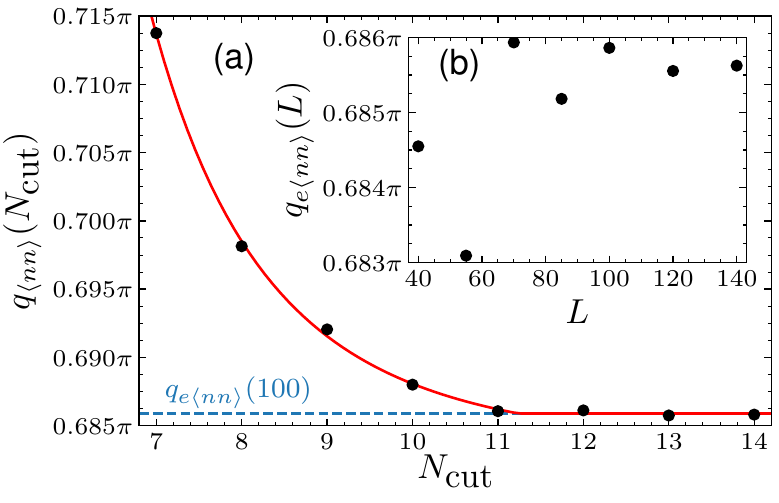}
  \caption{\label{fig:qe} (color online) (a) Position of $S(q)$ peak, $q_{\braket{nn}}$ (see Fig. \ref{fig:qm}) calculated for different values of maximum particles per site cutoff $N_{\textrm{cut}}$ for $L=100$. Red, solid line shows power-law decay to a constant value (reached for a finite $N_{\textrm{cut}}$, here approx. $11.24$), $q_{e\braket{nn}}$ (shown as a blue, dashed line).  (b) Values of $q_{e\braket{nn}}$ calculated for different system sizes ($L$). Data for (a) and (b) were calculated for $\mu/U=5.2$ and $t/U=0.6$. The position of $S(q)$ peak for $N_{\textrm{\small cut}}\geq 10$  $S(q)$ and for $L\geq 40$ changes by at least one order of magnitude less than the corresponding FWHM.}
\end{figure}

Some of the calculations (determination of boundaries of IPSS phase in Section \ref{sec:rhv}) were performed using a~smooth boundary DMRG method, referred to as a sine-squared deformation (SSD) DMRG. In this approach the Hamiltonian is rescaled using a~sine-squared deformation \cite{Hikihara2011}: $H_{\textrm{SDD}} = \sum_{j = 0}^{2} \sum_{i=1}^{L-j} f_{i, j} H_{i, i+j}$, where
\begin{equation}
  f_{i, j} = \sin^2\left[\frac{\pi}{L}\left(i + \frac{j-1}{2}\right)\right]
\end{equation} 
with $H_{i, i+j}$ acting only on sites $i$ and $i+j$ and $H_{i, i} \equiv H_i$ -- only on a~single site $i$.
We also add a~chemical potential term to the Hamiltonian, so that now $H_i = (U/2) n_i(n_i-1) - \mu n_i$. 

In contrast to regular DMRG methods, the density of the gas of particles (as measured in the middle part of the lattice) is not fixed by the number of particles $N$ used in the simulation, but rather by the value of $\mu$. An excess (or a deficit) of particles stemming from the choice of $N$ is taken care of by placing extra particles (vacancies) close to the system boundary, where the coefficient $f_{i,j}$ takes a minimal value.  This makes the edges act as an effective~bath for the particles (holes) in the middle of the lattice.
Because of that, in determination of the physical quantities, we consider only $40\%$ of the sites in the middle of the lattice, unless stated otherwise.

We pick $N$ such that it does not differ much from $L\rho_{\textrm{bulk}}$, the number compatible with the bulk density. This assures that fewer particles are displaced to (from) the edges, which minimizes the undesired boundary effects on the computed expectation values in the middle of the system.

In order to find the values of wavenumbers $q_{\braket{b^{\dag}b}}$ and $q_{\braket{nn}}$ [later plotted in Fig. \hyperref[kkrho_fig]{\ref*{kkrho_fig}(a)}] we look at the position of the peak of $S(q)$ (\ref{eq:sfac}) (or an analogical quantity for $\braket{b^{\dag}b}$ correlations).
To remove the boundary effects from our analysis, we only consider $m$ lattice sites in the middle when calculating the structure factor.
Depending on $m$, the position of the peak ($q_{m\braket{nn}}$) oscillates with decreasing amplitude (see Fig.~\ref{fig:qm}) around a value $q_{\braket{nn}}$, which is the one used in the main text.

As the mean densities in the IPSS phase in our calculations are quite high [with sites filled by more than 6 particles - Fig. \hyperref[kkrho_fig]{\ref*{kkrho_fig}(b)}], we calculated how the cutoff on maximum particles per site  ($N_{\textrm{cut}}$) in DMRG calculations affects the obtained value of $q_{\braket{nn}}$ [Fig. \hyperref[fig:qe]{\ref{fig:qe}(a)}], taking as an example values of $\mu/U = 5.2$ and $t/U = 0.6$, corresponding to $\rho_{\textrm{bulk}} \approx 3.1$.
For each system size $L$ we define $q_{e\braket{nn}}(L) = \lim_{N_{\textrm{cut}}\to\infty} q_{\braket{nn}}(N_{\textrm{cut}})$ and plot its value in Fig. \hyperref[fig:qe]{\ref{fig:qe}(b)}.
We determine that $N_{\textrm{cut}} = 12$ and $L=100$ are enough to get converged values of wavenumbers and these parameters were used for SSD DMRG calculations.

\bibliographystyle{apsrev4-1}
\bibliography{paper}

\begin{thebibliography}{60}%
\makeatletter
\providecommand \@ifxundefined [1]{%
 \@ifx{#1\undefined}
}%
\providecommand \@ifnum [1]{%
 \ifnum #1\expandafter \@firstoftwo
 \else \expandafter \@secondoftwo
 \fi
}%
\providecommand \@ifx [1]{%
 \ifx #1\expandafter \@firstoftwo
 \else \expandafter \@secondoftwo
 \fi
}%
\providecommand \natexlab [1]{#1}%
\providecommand \enquote  [1]{``#1''}%
\providecommand \bibnamefont  [1]{#1}%
\providecommand \bibfnamefont [1]{#1}%
\providecommand \citenamefont [1]{#1}%
\providecommand \href@noop [0]{\@secondoftwo}%
\providecommand \href [0]{\begingroup \@sanitize@url \@href}%
\providecommand \@href[1]{\@@startlink{#1}\@@href}%
\providecommand \@@href[1]{\endgroup#1\@@endlink}%
\providecommand \@sanitize@url [0]{\catcode `\\12\catcode `\$12\catcode
  `\&12\catcode `\#12\catcode `\^12\catcode `\_12\catcode `\%12\relax}%
\providecommand \@@startlink[1]{}%
\providecommand \@@endlink[0]{}%
\providecommand \url  [0]{\begingroup\@sanitize@url \@url }%
\providecommand \@url [1]{\endgroup\@href {#1}{\urlprefix }}%
\providecommand \urlprefix  [0]{URL }%
\providecommand \Eprint [0]{\href }%
\providecommand \doibase [0]{http://dx.doi.org/}%
\providecommand \selectlanguage [0]{\@gobble}%
\providecommand \bibinfo  [0]{\@secondoftwo}%
\providecommand \bibfield  [0]{\@secondoftwo}%
\providecommand \translation [1]{[#1]}%
\providecommand \BibitemOpen [0]{}%
\providecommand \bibitemStop [0]{}%
\providecommand \bibitemNoStop [0]{.\EOS\space}%
\providecommand \EOS [0]{\spacefactor3000\relax}%
\providecommand \BibitemShut  [1]{\csname bibitem#1\endcsname}%
\let\auto@bib@innerbib\@empty
\bibitem [{\citenamefont {Jaksch}\ \emph {et~al.}(1998)\citenamefont {Jaksch},
  \citenamefont {Bruder}, \citenamefont {Cirac}, \citenamefont {Gardiner},\
  and\ \citenamefont {Zoller}}]{Jaksch1998}%
  \BibitemOpen
  \bibfield  {author} {\bibinfo {author} {\bibfnamefont {D.}~\bibnamefont
  {Jaksch}}, \bibinfo {author} {\bibfnamefont {C.}~\bibnamefont {Bruder}},
  \bibinfo {author} {\bibfnamefont {J.~I.}\ \bibnamefont {Cirac}}, \bibinfo
  {author} {\bibfnamefont {C.~W.}\ \bibnamefont {Gardiner}}, \ and\ \bibinfo
  {author} {\bibfnamefont {P.}~\bibnamefont {Zoller}},\ }\href@noop {}
  {\bibfield  {journal} {\bibinfo  {journal} {Phys. Rev. Lett.}\ }\textbf
  {\bibinfo {volume} {81}},\ \bibinfo {pages} {3108} (\bibinfo {year}
  {1998})}\BibitemShut {NoStop}%
\bibitem [{\citenamefont {Greiner}\ \emph {et~al.}(2002)\citenamefont
  {Greiner}, \citenamefont {Mandel}, \citenamefont {Esslinger}, \citenamefont
  {H{\"a}nsch},\ and\ \citenamefont {Bloch}}]{Greiner2002}%
  \BibitemOpen
  \bibfield  {author} {\bibinfo {author} {\bibfnamefont {M.}~\bibnamefont
  {Greiner}}, \bibinfo {author} {\bibfnamefont {O.}~\bibnamefont {Mandel}},
  \bibinfo {author} {\bibfnamefont {T.}~\bibnamefont {Esslinger}}, \bibinfo
  {author} {\bibfnamefont {T.~W.}\ \bibnamefont {H{\"a}nsch}}, \ and\ \bibinfo
  {author} {\bibfnamefont {I.}~\bibnamefont {Bloch}},\ }\href@noop {}
  {\bibfield  {journal} {\bibinfo  {journal} {Nature}\ }\textbf {\bibinfo
  {volume} {415}},\ \bibinfo {pages} {39} (\bibinfo {year} {2002})}\BibitemShut
  {NoStop}%
\bibitem [{\citenamefont {Chin}\ \emph {et~al.}(2010)\citenamefont {Chin},
  \citenamefont {Grimm}, \citenamefont {Julienne},\ and\ \citenamefont
  {Tiesinga}}]{Chin2010}%
  \BibitemOpen
  \bibfield  {author} {\bibinfo {author} {\bibfnamefont {C.}~\bibnamefont
  {Chin}}, \bibinfo {author} {\bibfnamefont {R.}~\bibnamefont {Grimm}},
  \bibinfo {author} {\bibfnamefont {P.}~\bibnamefont {Julienne}}, \ and\
  \bibinfo {author} {\bibfnamefont {E.}~\bibnamefont {Tiesinga}},\ }\href@noop
  {} {\bibfield  {journal} {\bibinfo  {journal} {Rev. Mod. Phys.}\ }\textbf
  {\bibinfo {volume} {82}},\ \bibinfo {pages} {1225} (\bibinfo {year}
  {2010})}\BibitemShut {NoStop}%
\bibitem [{\citenamefont {Frisch}\ \emph {et~al.}(2015)\citenamefont {Frisch},
  \citenamefont {Mark}, \citenamefont {Aikawa}, \citenamefont {Baier},
  \citenamefont {Grimm}, \citenamefont {Petrov}, \citenamefont {Kotochigova},
  \citenamefont {Qu\'em\'ener}, \citenamefont {Lepers}, \citenamefont
  {Dulieu},\ and\ \citenamefont {Ferlaino}}]{Frisch2015}%
  \BibitemOpen
  \bibfield  {author} {\bibinfo {author} {\bibfnamefont {A.}~\bibnamefont
  {Frisch}}, \bibinfo {author} {\bibfnamefont {M.}~\bibnamefont {Mark}},
  \bibinfo {author} {\bibfnamefont {K.}~\bibnamefont {Aikawa}}, \bibinfo
  {author} {\bibfnamefont {S.}~\bibnamefont {Baier}}, \bibinfo {author}
  {\bibfnamefont {R.}~\bibnamefont {Grimm}}, \bibinfo {author} {\bibfnamefont
  {A.}~\bibnamefont {Petrov}}, \bibinfo {author} {\bibfnamefont
  {S.}~\bibnamefont {Kotochigova}}, \bibinfo {author} {\bibfnamefont
  {G.}~\bibnamefont {Qu\'em\'ener}}, \bibinfo {author} {\bibfnamefont
  {M.}~\bibnamefont {Lepers}}, \bibinfo {author} {\bibfnamefont
  {O.}~\bibnamefont {Dulieu}}, \ and\ \bibinfo {author} {\bibfnamefont
  {F.}~\bibnamefont {Ferlaino}},\ }\href {\doibase
  10.1103/PhysRevLett.115.203201} {\bibfield  {journal} {\bibinfo  {journal}
  {Phys. Rev. Lett.}\ }\textbf {\bibinfo {volume} {115}},\ \bibinfo {pages}
  {203201} (\bibinfo {year} {2015})}\BibitemShut {NoStop}%
\bibitem [{\citenamefont {Dalla~Torre}\ \emph {et~al.}(2006)\citenamefont
  {Dalla~Torre}, \citenamefont {Berg},\ and\ \citenamefont
  {Altman}}]{DallaTorre06}%
  \BibitemOpen
  \bibfield  {author} {\bibinfo {author} {\bibfnamefont {E.~G.}\ \bibnamefont
  {Dalla~Torre}}, \bibinfo {author} {\bibfnamefont {E.}~\bibnamefont {Berg}}, \
  and\ \bibinfo {author} {\bibfnamefont {E.}~\bibnamefont {Altman}},\ }\href
  {\doibase 10.1103/PhysRevLett.97.260401} {\bibfield  {journal} {\bibinfo
  {journal} {Phys. Rev. Lett.}\ }\textbf {\bibinfo {volume} {97}},\ \bibinfo
  {pages} {260401} (\bibinfo {year} {2006})}\BibitemShut {NoStop}%
\bibitem [{\citenamefont {Sinha}\ and\ \citenamefont
  {Santos}(2007)}]{Sinha2007}%
  \BibitemOpen
  \bibfield  {author} {\bibinfo {author} {\bibfnamefont {S.}~\bibnamefont
  {Sinha}}\ and\ \bibinfo {author} {\bibfnamefont {L.}~\bibnamefont {Santos}},\
  }\href {\doibase 10.1103/PhysRevLett.99.140406} {\bibfield  {journal}
  {\bibinfo  {journal} {Phys. Rev. Lett.}\ }\textbf {\bibinfo {volume} {99}},\
  \bibinfo {pages} {140406} (\bibinfo {year} {2007})}\BibitemShut {NoStop}%
\bibitem [{\citenamefont {Lahaye}\ \emph {et~al.}(2009)\citenamefont {Lahaye},
  \citenamefont {Menotti}, \citenamefont {Santos}, \citenamefont {Lewenstein},\
  and\ \citenamefont {Pfau}}]{Lahaye2009}%
  \BibitemOpen
  \bibfield  {author} {\bibinfo {author} {\bibfnamefont {T.}~\bibnamefont
  {Lahaye}}, \bibinfo {author} {\bibfnamefont {C.}~\bibnamefont {Menotti}},
  \bibinfo {author} {\bibfnamefont {L.}~\bibnamefont {Santos}}, \bibinfo
  {author} {\bibfnamefont {M.}~\bibnamefont {Lewenstein}}, \ and\ \bibinfo
  {author} {\bibfnamefont {T.}~\bibnamefont {Pfau}},\ }\href {\doibase
  10.1088/0034-4885/72/12/126401} {\bibfield  {journal} {\bibinfo  {journal}
  {Rep. Prog. Phys.}\ }\textbf {\bibinfo {volume} {72}},\ \bibinfo {pages}
  {126401} (\bibinfo {year} {2009})}\BibitemShut {NoStop}%
\bibitem [{\citenamefont {Ruhman}\ \emph {et~al.}(2012)\citenamefont {Ruhman},
  \citenamefont {Dalla~Torre}, \citenamefont {Huber},\ and\ \citenamefont
  {Altman}}]{Ruhman12}%
  \BibitemOpen
  \bibfield  {author} {\bibinfo {author} {\bibfnamefont {J.}~\bibnamefont
  {Ruhman}}, \bibinfo {author} {\bibfnamefont {E.~G.}\ \bibnamefont
  {Dalla~Torre}}, \bibinfo {author} {\bibfnamefont {S.~D.}\ \bibnamefont
  {Huber}}, \ and\ \bibinfo {author} {\bibfnamefont {E.}~\bibnamefont
  {Altman}},\ }\href {\doibase 10.1103/PhysRevB.85.125121} {\bibfield
  {journal} {\bibinfo  {journal} {Phys. Rev. B}\ }\textbf {\bibinfo {volume}
  {85}},\ \bibinfo {pages} {125121} (\bibinfo {year} {2012})}\BibitemShut
  {NoStop}%
\bibitem [{\citenamefont {Baranov}\ \emph {et~al.}(2012)\citenamefont
  {Baranov}, \citenamefont {Dalmonte}, \citenamefont {Pupillo},\ and\
  \citenamefont {Zoller}}]{Baranov2012}%
  \BibitemOpen
  \bibfield  {author} {\bibinfo {author} {\bibfnamefont {M.~A.}\ \bibnamefont
  {Baranov}}, \bibinfo {author} {\bibfnamefont {M.}~\bibnamefont {Dalmonte}},
  \bibinfo {author} {\bibfnamefont {G.}~\bibnamefont {Pupillo}}, \ and\
  \bibinfo {author} {\bibfnamefont {P.}~\bibnamefont {Zoller}},\ }\href
  {\doibase 10.1021/cr2003568} {\bibfield  {journal} {\bibinfo  {journal}
  {Chem. Rev.}\ }\textbf {\bibinfo {volume} {112}},\ \bibinfo {pages} {5012}
  (\bibinfo {year} {2012})},\ \bibinfo {note} {arXiv: 1207.1914}\BibitemShut
  {NoStop}%
\bibitem [{\citenamefont {Wall}\ and\ \citenamefont {Carr}(2013)}]{Wall2013}%
  \BibitemOpen
  \bibfield  {author} {\bibinfo {author} {\bibfnamefont {M.}~\bibnamefont
  {Wall}}\ and\ \bibinfo {author} {\bibfnamefont {L.}~\bibnamefont {Carr}},\
  }\href@noop {} {\bibfield  {journal} {\bibinfo  {journal} {New J. Phys.}\
  }\textbf {\bibinfo {volume} {15}},\ \bibinfo {pages} {123005} (\bibinfo
  {year} {2013})}\BibitemShut {NoStop}%
\bibitem [{\citenamefont {Gallem\'{\i}}\ \emph {et~al.}(2013)\citenamefont
  {Gallem\'{\i}}, \citenamefont {Guilleumas}, \citenamefont {Mayol},\ and\
  \citenamefont {Sanpera}}]{Gallemi13}%
  \BibitemOpen
  \bibfield  {author} {\bibinfo {author} {\bibfnamefont {A.}~\bibnamefont
  {Gallem\'{\i}}}, \bibinfo {author} {\bibfnamefont {M.}~\bibnamefont
  {Guilleumas}}, \bibinfo {author} {\bibfnamefont {R.}~\bibnamefont {Mayol}}, \
  and\ \bibinfo {author} {\bibfnamefont {A.}~\bibnamefont {Sanpera}},\ }\href
  {\doibase 10.1103/PhysRevA.88.063645} {\bibfield  {journal} {\bibinfo
  {journal} {Phys. Rev. A}\ }\textbf {\bibinfo {volume} {88}},\ \bibinfo
  {pages} {063645} (\bibinfo {year} {2013})}\BibitemShut {NoStop}%
\bibitem [{\citenamefont {Gammelmark}\ and\ \citenamefont
  {Zinner}(2013)}]{Gammelmark13}%
  \BibitemOpen
  \bibfield  {author} {\bibinfo {author} {\bibfnamefont {S.}~\bibnamefont
  {Gammelmark}}\ and\ \bibinfo {author} {\bibfnamefont {N.~T.}\ \bibnamefont
  {Zinner}},\ }\href {\doibase 10.1103/PhysRevB.88.245135} {\bibfield
  {journal} {\bibinfo  {journal} {Phys. Rev. B}\ }\textbf {\bibinfo {volume}
  {88}},\ \bibinfo {pages} {245135} (\bibinfo {year} {2013})}\BibitemShut
  {NoStop}%
\bibitem [{\citenamefont {Dutta}\ \emph {et~al.}(2015)\citenamefont {Dutta},
  \citenamefont {Gajda}, \citenamefont {Hauke}, \citenamefont {Lewenstein},
  \citenamefont {L{\"u}hmann}, \citenamefont {Malomed}, \citenamefont
  {Sowi{\'n}ski},\ and\ \citenamefont {Zakrzewski}}]{Dutta2015}%
  \BibitemOpen
  \bibfield  {author} {\bibinfo {author} {\bibfnamefont {O.}~\bibnamefont
  {Dutta}}, \bibinfo {author} {\bibfnamefont {M.}~\bibnamefont {Gajda}},
  \bibinfo {author} {\bibfnamefont {P.}~\bibnamefont {Hauke}}, \bibinfo
  {author} {\bibfnamefont {M.}~\bibnamefont {Lewenstein}}, \bibinfo {author}
  {\bibfnamefont {D.-S.}\ \bibnamefont {L{\"u}hmann}}, \bibinfo {author}
  {\bibfnamefont {B.~A.}\ \bibnamefont {Malomed}}, \bibinfo {author}
  {\bibfnamefont {T.}~\bibnamefont {Sowi{\'n}ski}}, \ and\ \bibinfo {author}
  {\bibfnamefont {J.}~\bibnamefont {Zakrzewski}},\ }\href@noop {} {\bibfield
  {journal} {\bibinfo  {journal} {Rep. Prog. Phys.}\ }\textbf {\bibinfo
  {volume} {78}},\ \bibinfo {pages} {066001} (\bibinfo {year}
  {2015})}\BibitemShut {NoStop}%
\bibitem [{\citenamefont {Gallem\'{\i}}\ \emph {et~al.}(2016)\citenamefont
  {Gallem\'{\i}}, \citenamefont {Queralt\'o}, \citenamefont {Guilleumas},
  \citenamefont {Mayol},\ and\ \citenamefont {Sanpera}}]{Gallemi16}%
  \BibitemOpen
  \bibfield  {author} {\bibinfo {author} {\bibfnamefont {A.}~\bibnamefont
  {Gallem\'{\i}}}, \bibinfo {author} {\bibfnamefont {G.}~\bibnamefont
  {Queralt\'o}}, \bibinfo {author} {\bibfnamefont {M.}~\bibnamefont
  {Guilleumas}}, \bibinfo {author} {\bibfnamefont {R.}~\bibnamefont {Mayol}}, \
  and\ \bibinfo {author} {\bibfnamefont {A.}~\bibnamefont {Sanpera}},\ }\href
  {\doibase 10.1103/PhysRevA.94.063626} {\bibfield  {journal} {\bibinfo
  {journal} {Phys. Rev. A}\ }\textbf {\bibinfo {volume} {94}},\ \bibinfo
  {pages} {063626} (\bibinfo {year} {2016})}\BibitemShut {NoStop}%
\bibitem [{\citenamefont {Kawaki}\ \emph {et~al.}(2017)\citenamefont {Kawaki},
  \citenamefont {Kuno},\ and\ \citenamefont {Ichinose}}]{Kawaki17}%
  \BibitemOpen
  \bibfield  {author} {\bibinfo {author} {\bibfnamefont {K.}~\bibnamefont
  {Kawaki}}, \bibinfo {author} {\bibfnamefont {Y.}~\bibnamefont {Kuno}}, \ and\
  \bibinfo {author} {\bibfnamefont {I.}~\bibnamefont {Ichinose}},\ }\href
  {\doibase 10.1103/PhysRevB.95.195101} {\bibfield  {journal} {\bibinfo
  {journal} {Phys. Rev. B}\ }\textbf {\bibinfo {volume} {95}},\ \bibinfo
  {pages} {195101} (\bibinfo {year} {2017})}\BibitemShut {NoStop}%
\bibitem [{\citenamefont {Cartarius}\ \emph {et~al.}(2017)\citenamefont
  {Cartarius}, \citenamefont {Minguzzi},\ and\ \citenamefont
  {Morigi}}]{Cartarius17}%
  \BibitemOpen
  \bibfield  {author} {\bibinfo {author} {\bibfnamefont {F.}~\bibnamefont
  {Cartarius}}, \bibinfo {author} {\bibfnamefont {A.}~\bibnamefont {Minguzzi}},
  \ and\ \bibinfo {author} {\bibfnamefont {G.}~\bibnamefont {Morigi}},\ }\href
  {\doibase 10.1103/PhysRevA.95.063603} {\bibfield  {journal} {\bibinfo
  {journal} {Phys. Rev. A}\ }\textbf {\bibinfo {volume} {95}},\ \bibinfo
  {pages} {063603} (\bibinfo {year} {2017})}\BibitemShut {NoStop}%
\bibitem [{\citenamefont {Baier}\ \emph {et~al.}(2016)\citenamefont {Baier},
  \citenamefont {Mark}, \citenamefont {Petter}, \citenamefont {Aikawa},
  \citenamefont {Chomaz}, \citenamefont {Cai}, \citenamefont {Baranov},
  \citenamefont {Zoller},\ and\ \citenamefont {Ferlaino}}]{Baier2016}%
  \BibitemOpen
  \bibfield  {author} {\bibinfo {author} {\bibfnamefont {S.}~\bibnamefont
  {Baier}}, \bibinfo {author} {\bibfnamefont {M.~J.}\ \bibnamefont {Mark}},
  \bibinfo {author} {\bibfnamefont {D.}~\bibnamefont {Petter}}, \bibinfo
  {author} {\bibfnamefont {K.}~\bibnamefont {Aikawa}}, \bibinfo {author}
  {\bibfnamefont {L.}~\bibnamefont {Chomaz}}, \bibinfo {author} {\bibfnamefont
  {Z.}~\bibnamefont {Cai}}, \bibinfo {author} {\bibfnamefont {M.}~\bibnamefont
  {Baranov}}, \bibinfo {author} {\bibfnamefont {P.}~\bibnamefont {Zoller}}, \
  and\ \bibinfo {author} {\bibfnamefont {F.}~\bibnamefont {Ferlaino}},\
  }\href@noop {} {\bibfield  {journal} {\bibinfo  {journal} {Science}\ }\textbf
  {\bibinfo {volume} {352}},\ \bibinfo {pages} {201} (\bibinfo {year}
  {2016})}\BibitemShut {NoStop}%
\bibitem [{\citenamefont {Delande}\ and\ \citenamefont
  {Zakrzewski}(2009)}]{Zakrzewski2009}%
  \BibitemOpen
  \bibfield  {author} {\bibinfo {author} {\bibfnamefont {D.}~\bibnamefont
  {Delande}}\ and\ \bibinfo {author} {\bibfnamefont {J.}~\bibnamefont
  {Zakrzewski}},\ }\href {\doibase 10.1103/PhysRevLett.102.085301} {\bibfield
  {journal} {\bibinfo  {journal} {Phys. Rev. Lett.}\ }\textbf {\bibinfo
  {volume} {102}},\ \bibinfo {pages} {085301} (\bibinfo {year}
  {2009})}\BibitemShut {NoStop}%
\bibitem [{\citenamefont {Gaunt}\ \emph {et~al.}(2013)\citenamefont {Gaunt},
  \citenamefont {Schmidutz}, \citenamefont {Gotlibovych}, \citenamefont
  {Smith},\ and\ \citenamefont {Hadzibabic}}]{Gaunt2013}%
  \BibitemOpen
  \bibfield  {author} {\bibinfo {author} {\bibfnamefont {A.~L.}\ \bibnamefont
  {Gaunt}}, \bibinfo {author} {\bibfnamefont {T.~F.}\ \bibnamefont
  {Schmidutz}}, \bibinfo {author} {\bibfnamefont {I.}~\bibnamefont
  {Gotlibovych}}, \bibinfo {author} {\bibfnamefont {R.~P.}\ \bibnamefont
  {Smith}}, \ and\ \bibinfo {author} {\bibfnamefont {Z.}~\bibnamefont
  {Hadzibabic}},\ }\href@noop {} {\bibfield  {journal} {\bibinfo  {journal}
  {Phys. Rev. Lett.}\ }\textbf {\bibinfo {volume} {110}},\ \bibinfo {pages}
  {200406} (\bibinfo {year} {2013})}\BibitemShut {NoStop}%
\bibitem [{\citenamefont {\L{\k{a}}cki}\ \emph {et~al.}(2016)\citenamefont
  {\L{\k{a}}cki}, \citenamefont {Pichler}, \citenamefont {Sterdyniak},
  \citenamefont {Lyras}, \citenamefont {Lembessis}, \citenamefont {Al-Dossary},
  \citenamefont {Budich},\ and\ \citenamefont {Zoller}}]{Lacki2016}%
  \BibitemOpen
  \bibfield  {author} {\bibinfo {author} {\bibfnamefont {M.}~\bibnamefont
  {\L{\k{a}}cki}}, \bibinfo {author} {\bibfnamefont {H.}~\bibnamefont
  {Pichler}}, \bibinfo {author} {\bibfnamefont {A.}~\bibnamefont {Sterdyniak}},
  \bibinfo {author} {\bibfnamefont {A.}~\bibnamefont {Lyras}}, \bibinfo
  {author} {\bibfnamefont {V.~E.}\ \bibnamefont {Lembessis}}, \bibinfo {author}
  {\bibfnamefont {O.}~\bibnamefont {Al-Dossary}}, \bibinfo {author}
  {\bibfnamefont {J.~C.}\ \bibnamefont {Budich}}, \ and\ \bibinfo {author}
  {\bibfnamefont {P.}~\bibnamefont {Zoller}},\ }\href@noop {} {\bibfield
  {journal} {\bibinfo  {journal} {Phys. Rev. A}\ }\textbf {\bibinfo {volume}
  {93}},\ \bibinfo {pages} {013604} (\bibinfo {year} {2016})}\BibitemShut
  {NoStop}%
\bibitem [{\citenamefont {Kim}\ \emph {et~al.}(2018)\citenamefont {Kim},
  \citenamefont {Zhu}, \citenamefont {Porto},\ and\ \citenamefont
  {Hafezi}}]{Kim2018}%
  \BibitemOpen
  \bibfield  {author} {\bibinfo {author} {\bibfnamefont {H.}~\bibnamefont
  {Kim}}, \bibinfo {author} {\bibfnamefont {G.}~\bibnamefont {Zhu}}, \bibinfo
  {author} {\bibfnamefont {J.}~\bibnamefont {Porto}}, \ and\ \bibinfo {author}
  {\bibfnamefont {M.}~\bibnamefont {Hafezi}},\ }\href@noop {} {\bibfield
  {journal} {\bibinfo  {journal} {arXiv preprint arXiv:1805.01483}\ } (\bibinfo
  {year} {2018})}\BibitemShut {NoStop}%
\bibitem [{\citenamefont {Schollw{\"o}ck}(2011)}]{Schollwock2011}%
  \BibitemOpen
  \bibfield  {author} {\bibinfo {author} {\bibfnamefont {U.}~\bibnamefont
  {Schollw{\"o}ck}},\ }\href@noop {} {\bibfield  {journal} {\bibinfo  {journal}
  {Ann. Phys. (N.Y.)}\ }\textbf {\bibinfo {volume} {326}},\ \bibinfo {pages}
  {96} (\bibinfo {year} {2011})}\BibitemShut {NoStop}%
\bibitem [{ITe()}]{ITensor}%
  \BibitemOpen
  \href@noop {} {\enquote {\bibinfo {title} {Itensor library},}\ }\bibinfo
  {howpublished} {\url{http://itensor.org}}\BibitemShut {NoStop}%
\bibitem [{\citenamefont {Rossini}\ and\ \citenamefont
  {Fazio}(2012)}]{Rossini2012}%
  \BibitemOpen
  \bibfield  {author} {\bibinfo {author} {\bibfnamefont {D.}~\bibnamefont
  {Rossini}}\ and\ \bibinfo {author} {\bibfnamefont {R.}~\bibnamefont
  {Fazio}},\ }\href@noop {} {\bibfield  {journal} {\bibinfo  {journal} {New J.
  Phys.}\ }\textbf {\bibinfo {volume} {14}},\ \bibinfo {pages} {065012}
  (\bibinfo {year} {2012})}\BibitemShut {NoStop}%
\bibitem [{\citenamefont {Batrouni}\ \emph {et~al.}(2014)\citenamefont
  {Batrouni}, \citenamefont {Rousseau}, \citenamefont {Scalettar},\ and\
  \citenamefont {Gr{\'e}maud}}]{Batrouni2014}%
  \BibitemOpen
  \bibfield  {author} {\bibinfo {author} {\bibfnamefont {G.}~\bibnamefont
  {Batrouni}}, \bibinfo {author} {\bibfnamefont {V.}~\bibnamefont {Rousseau}},
  \bibinfo {author} {\bibfnamefont {R.}~\bibnamefont {Scalettar}}, \ and\
  \bibinfo {author} {\bibfnamefont {B.}~\bibnamefont {Gr{\'e}maud}},\
  }\href@noop {} {\bibfield  {journal} {\bibinfo  {journal} {Phys. Rev. B}\
  }\textbf {\bibinfo {volume} {90}},\ \bibinfo {pages} {205123} (\bibinfo
  {year} {2014})}\BibitemShut {NoStop}%
\bibitem [{\citenamefont {Gr\'emaud}\ and\ \citenamefont
  {Batrouni}(2016)}]{Gremaud16}%
  \BibitemOpen
  \bibfield  {author} {\bibinfo {author} {\bibfnamefont {B.}~\bibnamefont
  {Gr\'emaud}}\ and\ \bibinfo {author} {\bibfnamefont {G.~G.}\ \bibnamefont
  {Batrouni}},\ }\href {\doibase 10.1103/PhysRevB.93.035108} {\bibfield
  {journal} {\bibinfo  {journal} {Phys. Rev. B}\ }\textbf {\bibinfo {volume}
  {93}},\ \bibinfo {pages} {035108} (\bibinfo {year} {2016})}\BibitemShut
  {NoStop}%
\bibitem [{\citenamefont {Wikberg}\ \emph {et~al.}(2012)\citenamefont
  {Wikberg}, \citenamefont {Larson}, \citenamefont {Bergholtz},\ and\
  \citenamefont {Karlhede}}]{Wikberg12}%
  \BibitemOpen
  \bibfield  {author} {\bibinfo {author} {\bibfnamefont {E.}~\bibnamefont
  {Wikberg}}, \bibinfo {author} {\bibfnamefont {J.}~\bibnamefont {Larson}},
  \bibinfo {author} {\bibfnamefont {E.~J.}\ \bibnamefont {Bergholtz}}, \ and\
  \bibinfo {author} {\bibfnamefont {A.}~\bibnamefont {Karlhede}},\ }\href
  {\doibase 10.1103/PhysRevA.85.033607} {\bibfield  {journal} {\bibinfo
  {journal} {Phys. Rev. A}\ }\textbf {\bibinfo {volume} {85}},\ \bibinfo
  {pages} {033607} (\bibinfo {year} {2012})}\BibitemShut {NoStop}%
\bibitem [{\citenamefont {\DJ{}uri\ifmmode~\acute{c}\else \'{c}\fi{}}\ \emph
  {et~al.}(2017)\citenamefont {\DJ{}uri\ifmmode~\acute{c}\else \'{c}\fi{}},
  \citenamefont {Biedro\ifmmode~\acute{n}\else \'{n}\fi{}},\ and\ \citenamefont
  {Zakrzewski}}]{Duric17}%
  \BibitemOpen
  \bibfield  {author} {\bibinfo {author} {\bibfnamefont {T.}~\bibnamefont
  {\DJ{}uri\ifmmode~\acute{c}\else \'{c}\fi{}}}, \bibinfo {author}
  {\bibfnamefont {K.}~\bibnamefont {Biedro\ifmmode~\acute{n}\else \'{n}\fi{}}},
  \ and\ \bibinfo {author} {\bibfnamefont {J.}~\bibnamefont {Zakrzewski}},\
  }\href {\doibase 10.1103/PhysRevB.95.085102} {\bibfield  {journal} {\bibinfo
  {journal} {Phys. Rev. B}\ }\textbf {\bibinfo {volume} {95}},\ \bibinfo
  {pages} {085102} (\bibinfo {year} {2017})}\BibitemShut {NoStop}%
\bibitem [{\citenamefont {Lu}\ \emph {et~al.}(2011)\citenamefont {Lu},
  \citenamefont {Burdick}, \citenamefont {Youn},\ and\ \citenamefont
  {Lev}}]{Lu2011}%
  \BibitemOpen
  \bibfield  {author} {\bibinfo {author} {\bibfnamefont {M.}~\bibnamefont
  {Lu}}, \bibinfo {author} {\bibfnamefont {N.~Q.}\ \bibnamefont {Burdick}},
  \bibinfo {author} {\bibfnamefont {S.~H.}\ \bibnamefont {Youn}}, \ and\
  \bibinfo {author} {\bibfnamefont {B.~L.}\ \bibnamefont {Lev}},\ }\href@noop
  {} {\bibfield  {journal} {\bibinfo  {journal} {Phys. Rev. Lett.}\ }\textbf
  {\bibinfo {volume} {107}},\ \bibinfo {pages} {190401} (\bibinfo {year}
  {2011})}\BibitemShut {NoStop}%
\bibitem [{\citenamefont {Aikawa}\ \emph {et~al.}(2012)\citenamefont {Aikawa},
  \citenamefont {Frisch}, \citenamefont {Mark}, \citenamefont {Baier},
  \citenamefont {Rietzler}, \citenamefont {Grimm},\ and\ \citenamefont
  {Ferlaino}}]{Aikawa2012}%
  \BibitemOpen
  \bibfield  {author} {\bibinfo {author} {\bibfnamefont {K.}~\bibnamefont
  {Aikawa}}, \bibinfo {author} {\bibfnamefont {A.}~\bibnamefont {Frisch}},
  \bibinfo {author} {\bibfnamefont {M.}~\bibnamefont {Mark}}, \bibinfo {author}
  {\bibfnamefont {S.}~\bibnamefont {Baier}}, \bibinfo {author} {\bibfnamefont
  {A.}~\bibnamefont {Rietzler}}, \bibinfo {author} {\bibfnamefont
  {R.}~\bibnamefont {Grimm}}, \ and\ \bibinfo {author} {\bibfnamefont
  {F.}~\bibnamefont {Ferlaino}},\ }\href@noop {} {\bibfield  {journal}
  {\bibinfo  {journal} {Phys. Rev. Lett.}\ }\textbf {\bibinfo {volume} {108}},\
  \bibinfo {pages} {210401} (\bibinfo {year} {2012})}\BibitemShut {NoStop}%
\bibitem [{\citenamefont {Chotia}\ \emph {et~al.}(2012)\citenamefont {Chotia},
  \citenamefont {Neyenhuis}, \citenamefont {Moses}, \citenamefont {Yan},
  \citenamefont {Covey}, \citenamefont {Foss-Feig}, \citenamefont {Rey},
  \citenamefont {Jin},\ and\ \citenamefont {Ye}}]{Chotia2012}%
  \BibitemOpen
  \bibfield  {author} {\bibinfo {author} {\bibfnamefont {A.}~\bibnamefont
  {Chotia}}, \bibinfo {author} {\bibfnamefont {B.}~\bibnamefont {Neyenhuis}},
  \bibinfo {author} {\bibfnamefont {S.~A.}\ \bibnamefont {Moses}}, \bibinfo
  {author} {\bibfnamefont {B.}~\bibnamefont {Yan}}, \bibinfo {author}
  {\bibfnamefont {J.~P.}\ \bibnamefont {Covey}}, \bibinfo {author}
  {\bibfnamefont {M.}~\bibnamefont {Foss-Feig}}, \bibinfo {author}
  {\bibfnamefont {A.~M.}\ \bibnamefont {Rey}}, \bibinfo {author} {\bibfnamefont
  {D.~S.}\ \bibnamefont {Jin}}, \ and\ \bibinfo {author} {\bibfnamefont
  {J.}~\bibnamefont {Ye}},\ }\href {\doibase 10.1103/PhysRevLett.108.080405}
  {\bibfield  {journal} {\bibinfo  {journal} {Phys. Rev. Lett.}\ }\textbf
  {\bibinfo {volume} {108}},\ \bibinfo {pages} {080405} (\bibinfo {year}
  {2012})}\BibitemShut {NoStop}%
\bibitem [{\citenamefont {Sowi\ifmmode~\acute{n}\else \'{n}\fi{}ski}\ \emph
  {et~al.}(2012)\citenamefont {Sowi\ifmmode~\acute{n}\else \'{n}\fi{}ski},
  \citenamefont {Dutta}, \citenamefont {Hauke}, \citenamefont {Tagliacozzo},\
  and\ \citenamefont {Lewenstein}}]{Sowinski2012}%
  \BibitemOpen
  \bibfield  {author} {\bibinfo {author} {\bibfnamefont {T.}~\bibnamefont
  {Sowi\ifmmode~\acute{n}\else \'{n}\fi{}ski}}, \bibinfo {author}
  {\bibfnamefont {O.}~\bibnamefont {Dutta}}, \bibinfo {author} {\bibfnamefont
  {P.}~\bibnamefont {Hauke}}, \bibinfo {author} {\bibfnamefont
  {L.}~\bibnamefont {Tagliacozzo}}, \ and\ \bibinfo {author} {\bibfnamefont
  {M.}~\bibnamefont {Lewenstein}},\ }\href {\doibase
  10.1103/PhysRevLett.108.115301} {\bibfield  {journal} {\bibinfo  {journal}
  {Phys. Rev. Lett.}\ }\textbf {\bibinfo {volume} {108}},\ \bibinfo {pages}
  {115301} (\bibinfo {year} {2012})}\BibitemShut {NoStop}%
\bibitem [{\citenamefont {Moses}\ \emph {et~al.}(2015)\citenamefont {Moses},
  \citenamefont {Covey}, \citenamefont {Miecnikowski}, \citenamefont {Yan},
  \citenamefont {Gadway}, \citenamefont {Ye},\ and\ \citenamefont
  {Jin}}]{Moses2015}%
  \BibitemOpen
  \bibfield  {author} {\bibinfo {author} {\bibfnamefont {S.~A.}\ \bibnamefont
  {Moses}}, \bibinfo {author} {\bibfnamefont {J.~P.}\ \bibnamefont {Covey}},
  \bibinfo {author} {\bibfnamefont {M.~T.}\ \bibnamefont {Miecnikowski}},
  \bibinfo {author} {\bibfnamefont {B.}~\bibnamefont {Yan}}, \bibinfo {author}
  {\bibfnamefont {B.}~\bibnamefont {Gadway}}, \bibinfo {author} {\bibfnamefont
  {J.}~\bibnamefont {Ye}}, \ and\ \bibinfo {author} {\bibfnamefont {D.~S.}\
  \bibnamefont {Jin}},\ }\href@noop {} {\bibfield  {journal} {\bibinfo
  {journal} {Science}\ }\textbf {\bibinfo {volume} {350}},\ \bibinfo {pages}
  {659} (\bibinfo {year} {2015})}\BibitemShut {NoStop}%
\bibitem [{\citenamefont {Moses}\ \emph {et~al.}(2017)\citenamefont {Moses},
  \citenamefont {Covey}, \citenamefont {Miecnikowski}, \citenamefont {Jin},\
  and\ \citenamefont {Ye}}]{Moses2017}%
  \BibitemOpen
  \bibfield  {author} {\bibinfo {author} {\bibfnamefont {S.~A.}\ \bibnamefont
  {Moses}}, \bibinfo {author} {\bibfnamefont {J.~P.}\ \bibnamefont {Covey}},
  \bibinfo {author} {\bibfnamefont {M.~T.}\ \bibnamefont {Miecnikowski}},
  \bibinfo {author} {\bibfnamefont {D.~S.}\ \bibnamefont {Jin}}, \ and\
  \bibinfo {author} {\bibfnamefont {J.}~\bibnamefont {Ye}},\ }\href@noop {}
  {\bibfield  {journal} {\bibinfo  {journal} {Nat. Phys.}\ }\textbf {\bibinfo
  {volume} {13}},\ \bibinfo {pages} {13} (\bibinfo {year} {2017})}\BibitemShut
  {NoStop}%
\bibitem [{\citenamefont {Sengupta}\ \emph {et~al.}(2005)\citenamefont
  {Sengupta}, \citenamefont {Pryadko}, \citenamefont {Alet}, \citenamefont
  {Troyer},\ and\ \citenamefont {Schmid}}]{Sengupta2005}%
  \BibitemOpen
  \bibfield  {author} {\bibinfo {author} {\bibfnamefont {P.}~\bibnamefont
  {Sengupta}}, \bibinfo {author} {\bibfnamefont {L.~P.}\ \bibnamefont
  {Pryadko}}, \bibinfo {author} {\bibfnamefont {F.}~\bibnamefont {Alet}},
  \bibinfo {author} {\bibfnamefont {M.}~\bibnamefont {Troyer}}, \ and\ \bibinfo
  {author} {\bibfnamefont {G.}~\bibnamefont {Schmid}},\ }\href {\doibase
  10.1103/PhysRevLett.94.207202} {\bibfield  {journal} {\bibinfo  {journal}
  {Phys. Rev. Lett.}\ }\textbf {\bibinfo {volume} {94}},\ \bibinfo {pages}
  {207202} (\bibinfo {year} {2005})}\BibitemShut {NoStop}%
\bibitem [{\citenamefont {Batrouni}\ \emph {et~al.}(2006)\citenamefont
  {Batrouni}, \citenamefont {H\'ebert},\ and\ \citenamefont
  {Scalettar}}]{Batrouni2006}%
  \BibitemOpen
  \bibfield  {author} {\bibinfo {author} {\bibfnamefont {G.~G.}\ \bibnamefont
  {Batrouni}}, \bibinfo {author} {\bibfnamefont {F.}~\bibnamefont {H\'ebert}},
  \ and\ \bibinfo {author} {\bibfnamefont {R.~T.}\ \bibnamefont {Scalettar}},\
  }\href {\doibase 10.1103/PhysRevLett.97.087209} {\bibfield  {journal}
  {\bibinfo  {journal} {Phys. Rev. Lett.}\ }\textbf {\bibinfo {volume} {97}},\
  \bibinfo {pages} {087209} (\bibinfo {year} {2006})}\BibitemShut {NoStop}%
\bibitem [{\citenamefont {Mishra}\ \emph {et~al.}(2009)\citenamefont {Mishra},
  \citenamefont {Pai}, \citenamefont {Ramanan}, \citenamefont {Luthra},\ and\
  \citenamefont {Das}}]{Mishra2009}%
  \BibitemOpen
  \bibfield  {author} {\bibinfo {author} {\bibfnamefont {T.}~\bibnamefont
  {Mishra}}, \bibinfo {author} {\bibfnamefont {R.~V.}\ \bibnamefont {Pai}},
  \bibinfo {author} {\bibfnamefont {S.}~\bibnamefont {Ramanan}}, \bibinfo
  {author} {\bibfnamefont {M.~S.}\ \bibnamefont {Luthra}}, \ and\ \bibinfo
  {author} {\bibfnamefont {B.~P.}\ \bibnamefont {Das}},\ }\href {\doibase
  10.1103/PhysRevA.80.043614} {\bibfield  {journal} {\bibinfo  {journal} {Phys.
  Rev. A}\ }\textbf {\bibinfo {volume} {80}},\ \bibinfo {pages} {043614}
  (\bibinfo {year} {2009})}\BibitemShut {NoStop}%
\bibitem [{\citenamefont {Capogrosso-Sansone}\ \emph
  {et~al.}(2010)\citenamefont {Capogrosso-Sansone}, \citenamefont {Trefzger},
  \citenamefont {Lewenstein}, \citenamefont {Zoller},\ and\ \citenamefont
  {Pupillo}}]{Capogrosso-Sansone2010}%
  \BibitemOpen
  \bibfield  {author} {\bibinfo {author} {\bibfnamefont {B.}~\bibnamefont
  {Capogrosso-Sansone}}, \bibinfo {author} {\bibfnamefont {C.}~\bibnamefont
  {Trefzger}}, \bibinfo {author} {\bibfnamefont {M.}~\bibnamefont
  {Lewenstein}}, \bibinfo {author} {\bibfnamefont {P.}~\bibnamefont {Zoller}},
  \ and\ \bibinfo {author} {\bibfnamefont {G.}~\bibnamefont {Pupillo}},\ }\href
  {\doibase 10.1103/PhysRevLett.104.125301} {\bibfield  {journal} {\bibinfo
  {journal} {Phys. Rev. Lett.}\ }\textbf {\bibinfo {volume} {104}},\ \bibinfo
  {pages} {125301} (\bibinfo {year} {2010})}\BibitemShut {NoStop}%
\bibitem [{\citenamefont {Zaletel}\ \emph {et~al.}(2014)\citenamefont
  {Zaletel}, \citenamefont {Parameswaran}, \citenamefont {R{\"u}egg},\ and\
  \citenamefont {Altman}}]{Zaletel2014}%
  \BibitemOpen
  \bibfield  {author} {\bibinfo {author} {\bibfnamefont {M.~P.}\ \bibnamefont
  {Zaletel}}, \bibinfo {author} {\bibfnamefont {S.}~\bibnamefont
  {Parameswaran}}, \bibinfo {author} {\bibfnamefont {A.}~\bibnamefont
  {R{\"u}egg}}, \ and\ \bibinfo {author} {\bibfnamefont {E.}~\bibnamefont
  {Altman}},\ }\href@noop {} {\bibfield  {journal} {\bibinfo  {journal} {Phys.
  Rev. B}\ }\textbf {\bibinfo {volume} {89}},\ \bibinfo {pages} {155142}
  (\bibinfo {year} {2014})}\BibitemShut {NoStop}%
\bibitem [{\citenamefont {Lewenstein}\ \emph {et~al.}(2012)\citenamefont
  {Lewenstein}, \citenamefont {Sanpera},\ and\ \citenamefont
  {Ahufinger}}]{Lewenstein2012}%
  \BibitemOpen
  \bibfield  {author} {\bibinfo {author} {\bibfnamefont {M.}~\bibnamefont
  {Lewenstein}}, \bibinfo {author} {\bibfnamefont {A.}~\bibnamefont {Sanpera}},
  \ and\ \bibinfo {author} {\bibfnamefont {V.}~\bibnamefont {Ahufinger}},\
  }\href@noop {} {\emph {\bibinfo {title} {Ultracold Atoms in Optical Lattices:
  Simulating quantum many-body systems}}}\ (\bibinfo  {publisher} {Oxford
  University Press},\ \bibinfo {year} {2012})\BibitemShut {NoStop}%
\bibitem [{\citenamefont {Kohn}(1959)}]{Kohn1959}%
  \BibitemOpen
  \bibfield  {author} {\bibinfo {author} {\bibfnamefont {W.}~\bibnamefont
  {Kohn}},\ }\href@noop {} {\bibfield  {journal} {\bibinfo  {journal} {Phys.
  Rev.}\ }\textbf {\bibinfo {volume} {115}},\ \bibinfo {pages} {809} (\bibinfo
  {year} {1959})}\BibitemShut {NoStop}%
\bibitem [{\citenamefont {\L\k{a}cki}\ and\ \citenamefont
  {Zakrzewski}(2013)}]{Lacki2013a}%
  \BibitemOpen
  \bibfield  {author} {\bibinfo {author} {\bibfnamefont {M.}~\bibnamefont
  {\L\k{a}cki}}\ and\ \bibinfo {author} {\bibfnamefont {J.}~\bibnamefont
  {Zakrzewski}},\ }\href {\doibase 10.1103/PhysRevLett.110.065301} {\bibfield
  {journal} {\bibinfo  {journal} {Phys. Rev. Lett.}\ }\textbf {\bibinfo
  {volume} {110}},\ \bibinfo {pages} {065301} (\bibinfo {year}
  {2013})}\BibitemShut {NoStop}%
\bibitem [{\citenamefont {Pichler}\ \emph {et~al.}(2013)\citenamefont
  {Pichler}, \citenamefont {Schachenmayer}, \citenamefont {Daley},\ and\
  \citenamefont {Zoller}}]{Pichler2013}%
  \BibitemOpen
  \bibfield  {author} {\bibinfo {author} {\bibfnamefont {H.}~\bibnamefont
  {Pichler}}, \bibinfo {author} {\bibfnamefont {J.}~\bibnamefont
  {Schachenmayer}}, \bibinfo {author} {\bibfnamefont {A.~J.}\ \bibnamefont
  {Daley}}, \ and\ \bibinfo {author} {\bibfnamefont {P.}~\bibnamefont
  {Zoller}},\ }\href@noop {} {\bibfield  {journal} {\bibinfo  {journal} {Phys.
  Rev. A}\ }\textbf {\bibinfo {volume} {87}},\ \bibinfo {pages} {033606}
  (\bibinfo {year} {2013})}\BibitemShut {NoStop}%
\bibitem [{\citenamefont {Maik}\ \emph {et~al.}(2013)\citenamefont {Maik},
  \citenamefont {Hauke}, \citenamefont {Dutta}, \citenamefont {Lewenstein},\
  and\ \citenamefont {Zakrzewski}}]{Maik2013}%
  \BibitemOpen
  \bibfield  {author} {\bibinfo {author} {\bibfnamefont {M.}~\bibnamefont
  {Maik}}, \bibinfo {author} {\bibfnamefont {P.}~\bibnamefont {Hauke}},
  \bibinfo {author} {\bibfnamefont {O.}~\bibnamefont {Dutta}}, \bibinfo
  {author} {\bibfnamefont {M.}~\bibnamefont {Lewenstein}}, \ and\ \bibinfo
  {author} {\bibfnamefont {J.}~\bibnamefont {Zakrzewski}},\ }\href@noop {}
  {\bibfield  {journal} {\bibinfo  {journal} {New J. Phys.}\ }\textbf {\bibinfo
  {volume} {15}},\ \bibinfo {pages} {113041} (\bibinfo {year}
  {2013})}\BibitemShut {NoStop}%
\bibitem [{\citenamefont {L{\"u}hmann}\ \emph {et~al.}(2012)\citenamefont
  {L{\"u}hmann}, \citenamefont {J{\"u}rgensen},\ and\ \citenamefont
  {Sengstock}}]{Luhmann2012}%
  \BibitemOpen
  \bibfield  {author} {\bibinfo {author} {\bibfnamefont {D.-S.}\ \bibnamefont
  {L{\"u}hmann}}, \bibinfo {author} {\bibfnamefont {O.}~\bibnamefont
  {J{\"u}rgensen}}, \ and\ \bibinfo {author} {\bibfnamefont {K.}~\bibnamefont
  {Sengstock}},\ }\href@noop {} {\bibfield  {journal} {\bibinfo  {journal} {New
  J. Phys.}\ }\textbf {\bibinfo {volume} {14}},\ \bibinfo {pages} {033021}
  (\bibinfo {year} {2012})}\BibitemShut {NoStop}%
\bibitem [{\citenamefont {Bissbort}\ \emph {et~al.}(2012)\citenamefont
  {Bissbort}, \citenamefont {Deuretzbacher},\ and\ \citenamefont
  {Hofstetter}}]{Bissbort2012}%
  \BibitemOpen
  \bibfield  {author} {\bibinfo {author} {\bibfnamefont {U.}~\bibnamefont
  {Bissbort}}, \bibinfo {author} {\bibfnamefont {F.}~\bibnamefont
  {Deuretzbacher}}, \ and\ \bibinfo {author} {\bibfnamefont {W.}~\bibnamefont
  {Hofstetter}},\ }\href@noop {} {\bibfield  {journal} {\bibinfo  {journal}
  {Phys. Rev. A}\ }\textbf {\bibinfo {volume} {86}},\ \bibinfo {pages} {023617}
  (\bibinfo {year} {2012})}\BibitemShut {NoStop}%
\bibitem [{\citenamefont {\L\k{a}cki}\ \emph {et~al.}(2013)\citenamefont
  {\L\k{a}cki}, \citenamefont {Delande},\ and\ \citenamefont
  {Zakrzewski}}]{Lacki2013}%
  \BibitemOpen
  \bibfield  {author} {\bibinfo {author} {\bibfnamefont {M.}~\bibnamefont
  {\L\k{a}cki}}, \bibinfo {author} {\bibfnamefont {D.}~\bibnamefont {Delande}},
  \ and\ \bibinfo {author} {\bibfnamefont {J.}~\bibnamefont {Zakrzewski}},\
  }\href {\doibase 10.1088/1367-2630/15/1/013062} {\bibfield  {journal}
  {\bibinfo  {journal} {New J. Phys.}\ }\textbf {\bibinfo {volume} {15}},\
  \bibinfo {pages} {013062} (\bibinfo {year} {2013})}\BibitemShut {NoStop}%
\bibitem [{\citenamefont {Kollath}\ \emph {et~al.}(2010)\citenamefont
  {Kollath}, \citenamefont {Roux}, \citenamefont {Biroli},\ and\ \citenamefont
  {L{\"a}uchli}}]{Kollath2010}%
  \BibitemOpen
  \bibfield  {author} {\bibinfo {author} {\bibfnamefont {C.}~\bibnamefont
  {Kollath}}, \bibinfo {author} {\bibfnamefont {G.}~\bibnamefont {Roux}},
  \bibinfo {author} {\bibfnamefont {G.}~\bibnamefont {Biroli}}, \ and\ \bibinfo
  {author} {\bibfnamefont {A.~M.}\ \bibnamefont {L{\"a}uchli}},\ }\href@noop {}
  {\bibfield  {journal} {\bibinfo  {journal} {J. Stat. Mech. Theory Exp.}\
  }\textbf {\bibinfo {volume} {2010}},\ \bibinfo {pages} {P08011} (\bibinfo
  {year} {2010})}\BibitemShut {NoStop}%
\bibitem [{\citenamefont {Trotzky}\ \emph {et~al.}(2012)\citenamefont
  {Trotzky}, \citenamefont {Chen}, \citenamefont {Flesch}, \citenamefont
  {McCulloch}, \citenamefont {Schollw{\"o}ck}, \citenamefont {Eisert},\ and\
  \citenamefont {Bloch}}]{Trotzky2012}%
  \BibitemOpen
  \bibfield  {author} {\bibinfo {author} {\bibfnamefont {S.}~\bibnamefont
  {Trotzky}}, \bibinfo {author} {\bibfnamefont {Y.-A.}\ \bibnamefont {Chen}},
  \bibinfo {author} {\bibfnamefont {A.}~\bibnamefont {Flesch}}, \bibinfo
  {author} {\bibfnamefont {I.~P.}\ \bibnamefont {McCulloch}}, \bibinfo {author}
  {\bibfnamefont {U.}~\bibnamefont {Schollw{\"o}ck}}, \bibinfo {author}
  {\bibfnamefont {J.}~\bibnamefont {Eisert}}, \ and\ \bibinfo {author}
  {\bibfnamefont {I.}~\bibnamefont {Bloch}},\ }\href@noop {} {\bibfield
  {journal} {\bibinfo  {journal} {Nat. Phys.}\ }\textbf {\bibinfo {volume}
  {8}},\ \bibinfo {pages} {325} (\bibinfo {year} {2012})}\BibitemShut {NoStop}%
\bibitem [{\citenamefont {G\'oral}\ \emph {et~al.}(2002)\citenamefont
  {G\'oral}, \citenamefont {Santos},\ and\ \citenamefont
  {Lewenstein}}]{Lewenstein2002}%
  \BibitemOpen
  \bibfield  {author} {\bibinfo {author} {\bibfnamefont {K.}~\bibnamefont
  {G\'oral}}, \bibinfo {author} {\bibfnamefont {L.}~\bibnamefont {Santos}}, \
  and\ \bibinfo {author} {\bibfnamefont {M.}~\bibnamefont {Lewenstein}},\
  }\href {\doibase 10.1103/PhysRevLett.88.170406} {\bibfield  {journal}
  {\bibinfo  {journal} {Phys. Rev. Lett.}\ }\textbf {\bibinfo {volume} {88}},\
  \bibinfo {pages} {170406} (\bibinfo {year} {2002})}\BibitemShut {NoStop}%
\bibitem [{\citenamefont {Salger}\ \emph {et~al.}(2007)\citenamefont {Salger},
  \citenamefont {Geckeler}, \citenamefont {Kling},\ and\ \citenamefont
  {Weitz}}]{Salger2007}%
  \BibitemOpen
  \bibfield  {author} {\bibinfo {author} {\bibfnamefont {T.}~\bibnamefont
  {Salger}}, \bibinfo {author} {\bibfnamefont {C.}~\bibnamefont {Geckeler}},
  \bibinfo {author} {\bibfnamefont {S.}~\bibnamefont {Kling}}, \ and\ \bibinfo
  {author} {\bibfnamefont {M.}~\bibnamefont {Weitz}},\ }\href {\doibase
  10.1103/PhysRevLett.99.190405} {\bibfield  {journal} {\bibinfo  {journal}
  {Phys. Rev. Lett.}\ }\textbf {\bibinfo {volume} {99}},\ \bibinfo {pages}
  {190405} (\bibinfo {year} {2007})}\BibitemShut {NoStop}%
\bibitem [{\citenamefont {Yi}\ \emph {et~al.}(2008)\citenamefont {Yi},
  \citenamefont {Daley}, \citenamefont {Pupillo},\ and\ \citenamefont
  {Zoller}}]{Yi2008}%
  \BibitemOpen
  \bibfield  {author} {\bibinfo {author} {\bibfnamefont {W.}~\bibnamefont
  {Yi}}, \bibinfo {author} {\bibfnamefont {A.~J.}\ \bibnamefont {Daley}},
  \bibinfo {author} {\bibfnamefont {G.}~\bibnamefont {Pupillo}}, \ and\
  \bibinfo {author} {\bibfnamefont {P.}~\bibnamefont {Zoller}},\ }\href
  {http://stacks.iop.org/1367-2630/10/i=7/a=073015} {\bibfield  {journal}
  {\bibinfo  {journal} {New J. Phys.}\ }\textbf {\bibinfo {volume} {10}},\
  \bibinfo {pages} {073015} (\bibinfo {year} {2008})}\BibitemShut {NoStop}%
\bibitem [{\citenamefont {Sun}\ \emph {et~al.}(2011)\citenamefont {Sun},
  \citenamefont {Evers}, \citenamefont {Kiffner},\ and\ \citenamefont
  {Zubairy}}]{Sun2011}%
  \BibitemOpen
  \bibfield  {author} {\bibinfo {author} {\bibfnamefont {Q.}~\bibnamefont
  {Sun}}, \bibinfo {author} {\bibfnamefont {J.}~\bibnamefont {Evers}}, \bibinfo
  {author} {\bibfnamefont {M.}~\bibnamefont {Kiffner}}, \ and\ \bibinfo
  {author} {\bibfnamefont {M.~S.}\ \bibnamefont {Zubairy}},\ }\href {\doibase
  10.1103/PhysRevA.83.053412} {\bibfield  {journal} {\bibinfo  {journal} {Phys.
  Rev. A}\ }\textbf {\bibinfo {volume} {83}},\ \bibinfo {pages} {053412}
  (\bibinfo {year} {2011})}\BibitemShut {NoStop}%
\bibitem [{\citenamefont {Nascimbene}\ \emph {et~al.}(2015)\citenamefont
  {Nascimbene}, \citenamefont {Goldman}, \citenamefont {Cooper},\ and\
  \citenamefont {Dalibard}}]{Nascimbene2015}%
  \BibitemOpen
  \bibfield  {author} {\bibinfo {author} {\bibfnamefont {S.}~\bibnamefont
  {Nascimbene}}, \bibinfo {author} {\bibfnamefont {N.}~\bibnamefont {Goldman}},
  \bibinfo {author} {\bibfnamefont {N.~R.}\ \bibnamefont {Cooper}}, \ and\
  \bibinfo {author} {\bibfnamefont {J.}~\bibnamefont {Dalibard}},\ }\href
  {\doibase 10.1103/PhysRevLett.115.140401} {\bibfield  {journal} {\bibinfo
  {journal} {Phys. Rev. Lett.}\ }\textbf {\bibinfo {volume} {115}},\ \bibinfo
  {pages} {140401} (\bibinfo {year} {2015})}\BibitemShut {NoStop}%
\bibitem [{\citenamefont {Giamarchi}(2004)}]{giamarchi2004}%
  \BibitemOpen
  \bibfield  {author} {\bibinfo {author} {\bibfnamefont {T.}~\bibnamefont
  {Giamarchi}},\ }\href@noop {} {\emph {\bibinfo {title} {Quantum physics in
  one dimension}}},\ Vol.\ \bibinfo {volume} {121}\ (\bibinfo  {publisher}
  {Oxford university press},\ \bibinfo {year} {2004})\BibitemShut {NoStop}%
\bibitem [{\citenamefont {K{\"u}hner}\ \emph {et~al.}(2000)\citenamefont
  {K{\"u}hner}, \citenamefont {White},\ and\ \citenamefont
  {Monien}}]{Kuhner2000}%
  \BibitemOpen
  \bibfield  {author} {\bibinfo {author} {\bibfnamefont {T.~D.}\ \bibnamefont
  {K{\"u}hner}}, \bibinfo {author} {\bibfnamefont {S.~R.}\ \bibnamefont
  {White}}, \ and\ \bibinfo {author} {\bibfnamefont {H.}~\bibnamefont
  {Monien}},\ }\href@noop {} {\bibfield  {journal} {\bibinfo  {journal} {Phys.
  Rev. B}\ }\textbf {\bibinfo {volume} {61}},\ \bibinfo {pages} {12474}
  (\bibinfo {year} {2000})}\BibitemShut {NoStop}%
\bibitem [{\citenamefont {Jiang}\ \emph {et~al.}(2012)\citenamefont {Jiang},
  \citenamefont {Fu},\ and\ \citenamefont {Xu}}]{Jiang2012}%
  \BibitemOpen
  \bibfield  {author} {\bibinfo {author} {\bibfnamefont {H.-C.}\ \bibnamefont
  {Jiang}}, \bibinfo {author} {\bibfnamefont {L.}~\bibnamefont {Fu}}, \ and\
  \bibinfo {author} {\bibfnamefont {C.}~\bibnamefont {Xu}},\ }\href {\doibase
  10.1103/PhysRevB.86.045129} {\bibfield  {journal} {\bibinfo  {journal} {Phys.
  Rev. B}\ }\textbf {\bibinfo {volume} {86}},\ \bibinfo {pages} {045129}
  (\bibinfo {year} {2012})}\BibitemShut {NoStop}%
\bibitem [{\citenamefont {Rapp}\ \emph {et~al.}(2012)\citenamefont {Rapp},
  \citenamefont {Deng},\ and\ \citenamefont {Santos}}]{Rapp2012}%
  \BibitemOpen
  \bibfield  {author} {\bibinfo {author} {\bibfnamefont {A.}~\bibnamefont
  {Rapp}}, \bibinfo {author} {\bibfnamefont {X.}~\bibnamefont {Deng}}, \ and\
  \bibinfo {author} {\bibfnamefont {L.}~\bibnamefont {Santos}},\ }\href
  {\doibase 10.1103/PhysRevLett.109.203005} {\bibfield  {journal} {\bibinfo
  {journal} {Phys. Rev. Lett.}\ }\textbf {\bibinfo {volume} {109}},\ \bibinfo
  {pages} {203005} (\bibinfo {year} {2012})}\BibitemShut {NoStop}%
\bibitem [{\citenamefont {Wierschem}\ and\ \citenamefont
  {Sengupta}(2014)}]{Wierschem14}%
  \BibitemOpen
  \bibfield  {author} {\bibinfo {author} {\bibfnamefont {K.}~\bibnamefont
  {Wierschem}}\ and\ \bibinfo {author} {\bibfnamefont {P.}~\bibnamefont
  {Sengupta}},\ }\href@noop {} {\bibfield  {journal} {\bibinfo  {journal} {Mod.
  Phys. Lett. B}\ }\textbf {\bibinfo {volume} {28}},\ \bibinfo {pages}
  {1430017} (\bibinfo {year} {2014})}\BibitemShut {NoStop}%
\bibitem [{\citenamefont {Hikihara}\ and\ \citenamefont
  {Nishino}(2011)}]{Hikihara2011}%
  \BibitemOpen
  \bibfield  {author} {\bibinfo {author} {\bibfnamefont {T.}~\bibnamefont
  {Hikihara}}\ and\ \bibinfo {author} {\bibfnamefont {T.}~\bibnamefont
  {Nishino}},\ }\href {\doibase 10.1103/PhysRevB.83.060414} {\bibfield
  {journal} {\bibinfo  {journal} {Phys. Rev. B}\ }\textbf {\bibinfo {volume}
  {83}},\ \bibinfo {pages} {060414} (\bibinfo {year} {2011})}\BibitemShut
  {NoStop}%
\end{thebibliography}%

\end{document}